\documentclass[traditabstract]{aa}  
\newcommand{\hii}{H\,{\scriptsize II}}

\newcommand{\av}{A$_{\rm v}$}  
\usepackage{epsfig,graphicx,color}  
\usepackage{ctable}  
  
\usepackage{txfonts}  
\usepackage{ctable}  
  
\begin{document}

\title{Understanding star formation in molecular clouds}  
\subtitle{III. Probability distribution functions of molecular lines in Cygnus X}   
  
  \author{N.Schneider \inst{1,2}  
  \and S. Bontemps  \inst{1}   
  \and F. Motte   \inst{3}  
  \and V. Ossenkopf  \inst{2}  
  \and R.S. Klessen  \inst{4}  
  \and R. Simon    \inst{2}  
  \and S. Fechtenbaum \inst{1}  
  \and F. Herpin \inst{1}
  \and P. Tremblin     \inst{5}
  \and T. Csengeri \inst{6}
  \and P.C. Myers \inst{7}
  \and T. Hill     \inst{8}
  \and M. Cunningham \inst{9}
  \and C. Federrath     \inst{10}  
  }      
 \institute{Univ. Bordeaux, LAB, CNRS, UMR 5804, 33270, Floirac, France 
  \and   
  I.\,Physikalisches Institut, Universit\"at zu K\"oln,  
  Z\"ulpicher Stra{\ss}e 77,  50937 K\"oln, Germany
  \and   
  IRFU/SAp CEA/DSM, Laboratoire AIM CNRS - Universit\'e Paris   
  Diderot, 91191 Gif-sur-Yvette, France  
  \and   
  Universit\"at Heidelberg, Zentrum f\"ur Astronomie, Albert-Ueberle Str. 2, 69120 Heidelberg, Germany
  \and   
  Maison de la simulation, CEA Saclay, 91191 Gif-sur-Yvette, France 
  \and   
  Max-Planck Institut f\"ur Radioastronomie, Auf dem H\"ugel, Bonn, Germany   
  \and   
  Harvard-Smithsonian Center for Astrophysics, 60 Garden Street, Cambridge MA 02138, USA
  \and   
  Joint ALMA observatory, Santiago, Chile
  \and   
  School of Physics, University of New South Wales, Sydney, NSW 2052, Australia
  \and   
  Research School of Astronomy and Astrophysics, The Australian National University, Canberra, ACT 2611, Australia
}  
  
  
\mail{nschneid@ph1.uni-koeln.de}  
  
\titlerunning{III. molecular line PDFs}  
\authorrunning{N. Schneider}  
  
\date{\today}  
  
\date{Received August 10, 2015; accepted November 22, 2015}  
  
\abstract {The probability distribution function of column density
  (N-PDF) serves as a powerful tool to characterise the various physical
  processes that influence the structure of molecular clouds. Studies
  that use extinction maps or H$_2$ column-density maps ($N$) that
  are derived from dust show that star-forming clouds can best be
  characterised by lognormal PDFs for the lower $N$ range and a
  power-law tail for higher $N$, which is commonly attributed to
  turbulence and self-gravity and/or pressure, respectively.  While
  PDFs from dust cover a large dynamic range (typically $N
  \sim$10$^{20-24}$ cm$^{-2}$ or \av$\sim$0.1--1000), PDFs obtained
  from molecular lines -- converted into H$_2$ column density --
  potentially trace more selectively different regimes of (column)
  densities and temperatures. They also enable us to distinguish
  different clouds along the line of sight through using the velocity
  information. We report here on PDFs that were obtained from
  observations of $^{12}$CO, $^{13}$CO, C$^{18}$O, CS, and N$_2$H$^+$
  in the Cygnus X North region, and make a comparison to a PDF that
  was derived from dust observations with the {\sl Herschel}
  satellite. The PDF of $^{12}$CO is lognormal for \av$\sim$1--30, but
  is cut for higher \av\ because of optical depth effects.  The PDFs
  of C$^{18}$O and $^{13}$CO are mostly lognormal up to
  \av$\sim$1--15, followed by excess up to \av$\sim$40. Above that
  value, all CO PDFs drop, which is most likely due to depletion. The
  high density tracers CS and N$_2$H$^+$ exhibit only a power law
  distribution between \av$\sim$15 and 400, respectively. The PDF from
  dust is lognormal for \av$\sim$3--15 and has a power-law tail up to
  \av$\sim$500.  Absolute values for the molecular line column
  densities are, however, rather uncertain because of abundance and
  excitation temperature variations. If we take the dust PDF at face
  value, we `calibrate' the molecular line PDF of CS to that of the
  dust and determine an abundance [CS]/[H$_2$] of 10$^{-9}$.  The
  slopes of the power-law tails of the CS, N$_2$H$^+$, and dust PDFs
  are -1.6, -1.4, and -2.3, respectively, and are thus consistent with
  free-fall collapse of filaments and clumps. A quasi static
  configuration of filaments and clumps can also possibly account for
  the observed N-pdfs, providing they have a sufficiently condensed
  density structure and external ram pressure by gas accretion is
  provided. The somehow flatter slopes of N$_2$H$^+$ and CS can
  reflect an abundance change and/or subthermal excitation at low
  column densities.}

  
   \maketitle  
  

\begin{figure}[!htpb]  
\begin{centering}  
\includegraphics [width=8cm, angle={0}]{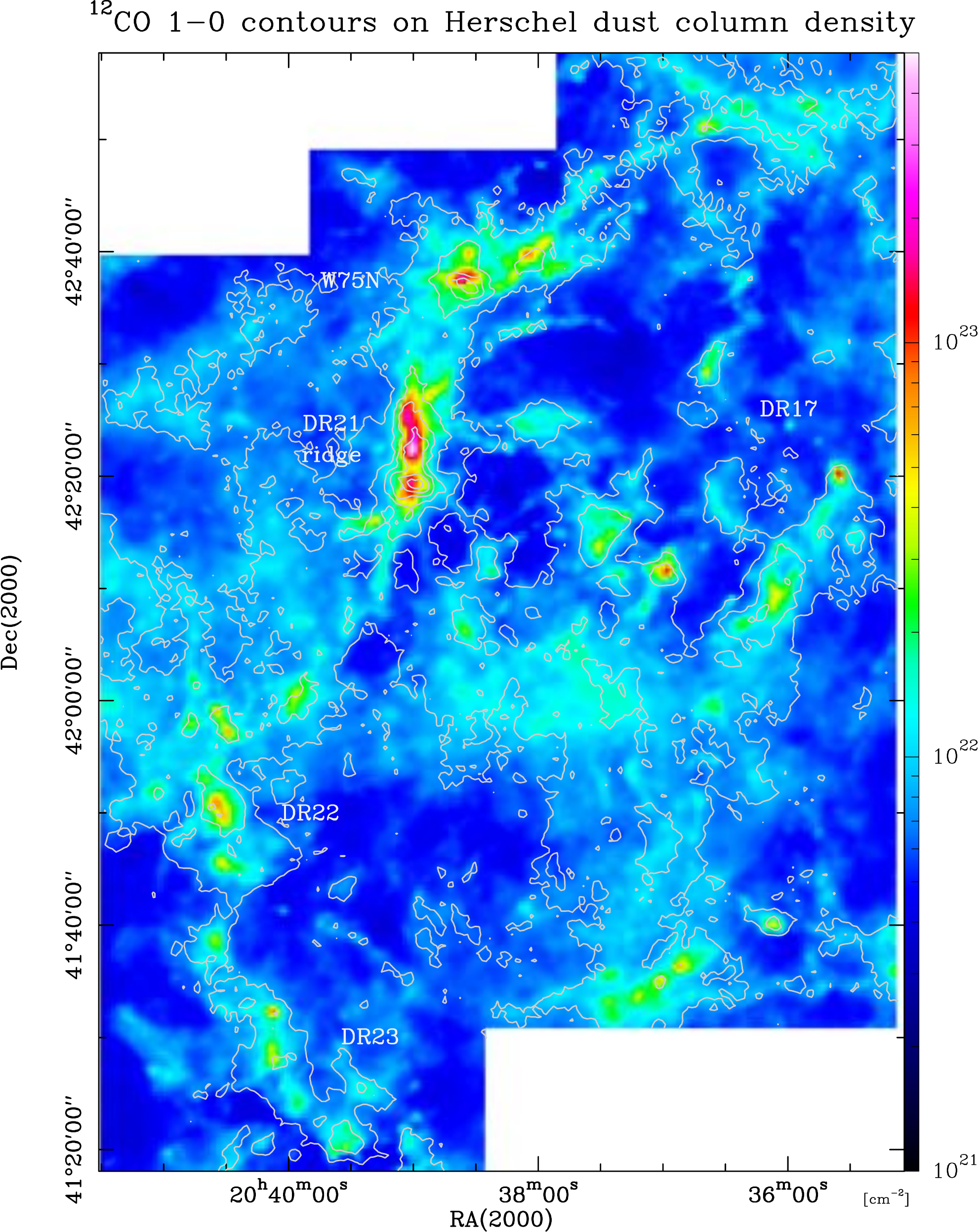}  
\caption[] {$^{12}$CO 1$\to$0 contours of line integrated (v=-10 to 20
  km s$^{-1}$) emission with levels 63.5 to 381 K kms$^{-1}$ in steps of
  63.5 K kms$^{-1}$ overlaid on a {\sl Herschel} dust column-density
  map at 36$''$ resolution (that corresponds to 0.41 pc at a distance
  of 1.4 kpc). This map was corrected for an average foreground
  contamination of \av = 5. Prominent features, such as W75N or the DR21
  ridge, are indicated in the plot.  DR17, 22, and 23 are \hii\ 
  regions.}
\label{herschel-co}  
\end{centering} 
\end{figure}  
  
\section{Introduction} \label{intro}  
 
Probability distribution functions (PDFs) form the basis of many
modern theories of star formation (e.g. Krumholz \& McKee
\cite{krumholz2005}; Hennebelle \& Chabrier \cite{hennebelle2008};
Federrath et al. \cite{fed2008}, \cite{fed2010}; Padoan \& Nordlund
\cite{padoan2011}; Federrath \& Klessen \cite{fed2012},
\cite{fed2013}; Hopkins \cite{hopkins2013}), and are frequently used
to characterise properties of the interstellar medium in simulations
(Klessen \cite{klessen2000}; Vazquez-Semadeni \& Garcia
\cite{vaz2001}; Burkhart et al. \cite{burk2013}, Ward et al.
\cite{ward2014}).  In summary, a PDF is defined as the probability of
finding gas within a column-density\footnote{The column density $N$
  can be expressed as visual extinction \av\, with N(H$_2$) =
  \av\ 0.94$\times$10$^{21}$ cm$^{-2}$ mag$^{-1}$ (Bohlin et al.
  \cite{bohlin1978}).} range [$N$, $N$+d$N$].  We define
$\eta\equiv\rm \ln(N/ \langle N \rangle)$, and the quantity
$p_\eta(\eta)$ then corresponds to the PDF of $\eta$.  For more
details on the basic definitions, see Schneider et
al. (\cite{schneider2015a}).

\begin{figure*}[!htpb]  
\begin{centering}  
\includegraphics [width=17cm, angle={0}]{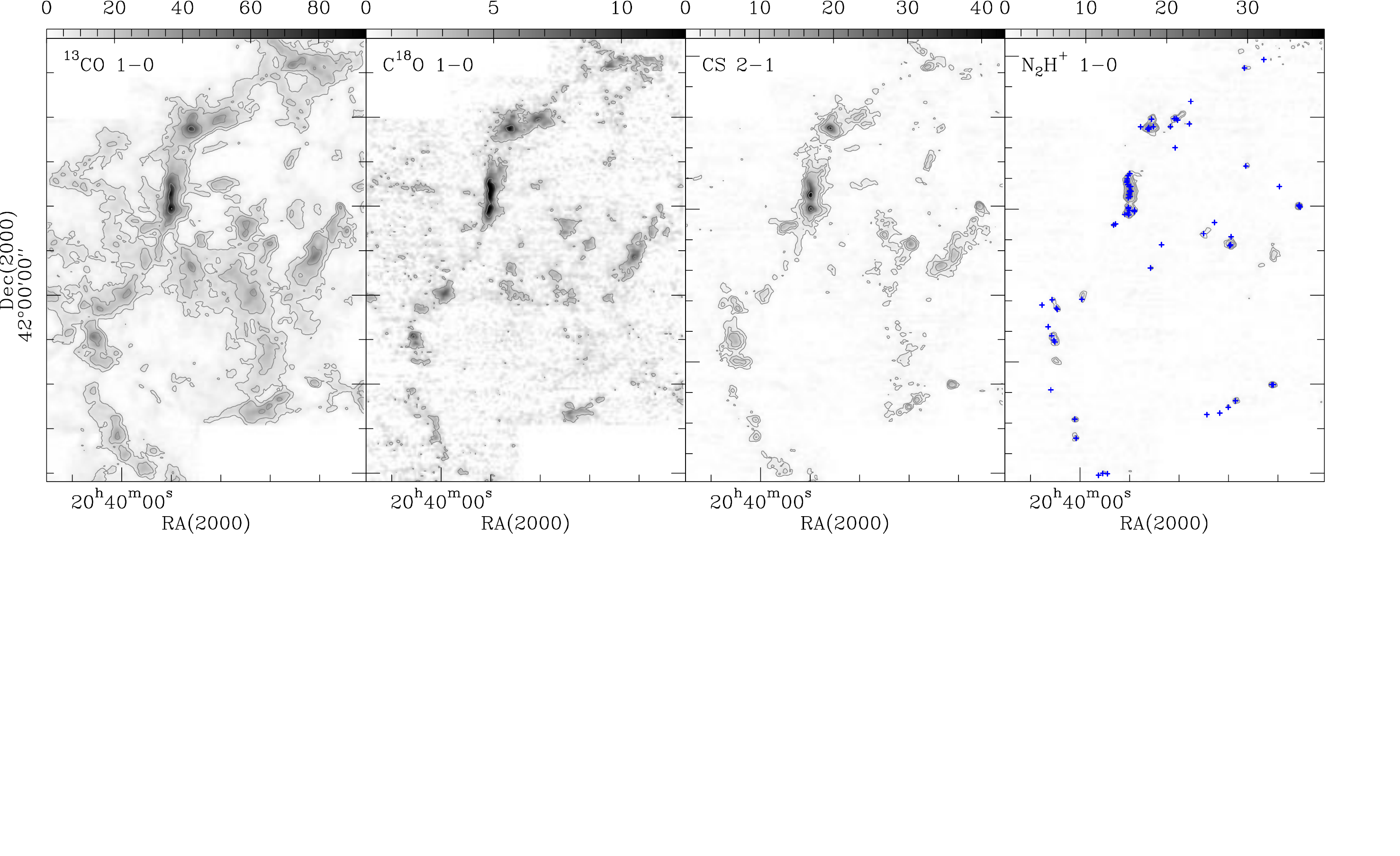}  
\vspace{-4cm}
\caption[] {Line integrated (v=-10 to 20 km s$^{-1}$)
  maps (in K kms$^{-1}$) of molecular emission in Cygnus X North. The
  lower level is zero for all maps. The blue crosses in the N$_2$H$^+$ map
  indicate submm-continuum sources (Motte et al. \cite{motte2007}).}
\label{maps}  
\end{centering} 
\end{figure*}

Observationally, PDFs were first obtained from near-IR extinction maps
(Lombardi et al. \cite{lombardi2008}, Kainulainen et al.
\cite{kai2009}, Froebrich \& Rowles \cite{froebrich2010}). With the
advent of {\sl Herschel}\footnote{Herschel is an ESA space observatory
  with science instruments provided by European-led Principal
  Investigator consortia and with important participation from
  NASA.}, it is now possible to determine dust column-density maps
that cover a very large dynamic range from \av$<$1 up to a few hundred
\av .  Studies using this type of map reveal large variations in dust
PDFs.  Earlier findings (Kainulainen et al. 2009) of a simple
lognormal plus power-law tail distribution for star-forming regions
are confirmed (e.g. Schneider et al. \cite{schneider2013},
\cite{schneider2015a}; Stutz \& Kainulainen \cite{stutz2015}), but
more complex shapes are also detected (Hill et al. \cite{hill2011};
Schneider et al.  \cite{schneider2012}, \cite{schneider2015c}; Russeil
et al. \cite{russeil2013}; Alves de Oliveira \cite{catarina2014};
Tremblin et al. \cite{tremblin2014}).  The outcome of these works is
that (i) all clouds have a distribution for low \av\, that is best
described by a lognormal, (ii) a second peak at low \av\ can emerge in
the case of external pressure, which represents the compressed shell,
and (iii) a single or double power-law tail is found for star-forming
regions, and it is proposed that this is related to gravitational
collapse and/or external pressure (either as a phase transition
between clumps and interclump gas or stellar feedback).

Only a few studies attempt to make PDFs from molecular line
observations. $N({\rm H_2})$-PDFs of $^{12}$CO or $^{13}$CO (Goldsmith
et al. \cite{goldsmith2008}, Wong et al. \cite{wong2008}, Goodman et
al.  \cite{good2009}, Lo et al. \cite{lo2009}, Carlhoff et
al. \cite{carlhoff2013}, Schneider et al.  \cite{schneider2015b}) are
often clipped at a certain \av\, threshold because the lines can
become optically thick and CO is depleted. Taking W43 as an example,
this threshold for $^{13}$CO 2$\to$1 lies between \av\ = 30 (for an
assumed T$_{ex}$=10 K) and $\sim$100 (for T$_{ex}$=12 K and correction
for high opacity). Assembling a PDF from one or several molecular line
tracers with high critical densities such as CS, HCO$^+$, HCN,
N$_2$H$^+$ etc. has not yet been tried owing to the lack of extended
maps. In this study, we use our large dataset of Cygnus X in the
molecular lines of $^{12}$CO, $^{13}$CO, and C$^{18}$O 1$\to$0, CS
2$\to$1, and N$_2$H$^+$ 1$\to$0 that were obtained with the Five
Colleges Radio Astronomy Observatory (FCRAO), and dust column-density
maps of the Cygnus X North region (Hennemann et al.
\cite{hennemann2012}, Schneider et al. \cite{schneider2015d}) to
produce PDFs from H$_2$ column density that were obtained from
molecules and dust. We discuss their properties but also point out the
large uncertainties and observational difficulties related to
molecular line PDFs.

\section{Observations} \label{obs}  

\noindent {\bf Molecular line data} \\
\noindent
We use data from a molecular line survey of $^{12}$CO (115.3 GHz),
$^{13}$CO (110.2 GHz), C$^{18}$O 1$\to$0 (109.8 GHz), CS 2$\to$1 (98.0
GHz), and N$_2$H$^+$ 1$\to$0 (93.2 GHz), which were collected with the
FCRAO 14m radiotelescope between 2003 and 2006.  The data were
obtained using the 32 pixel Second Quabbin Optical Imaging Array
(SEQUOIA) in an on-the-fly (OTF) observing mode. The beamwidth of the
FCRAO at 93 GHz is $\sim$48$''$ and at 110 GHz $\sim$45$''$, while the
main beam efficiency at that time was 0.48.  For the present paper, we
use maps on a 20$''$ grid.  The data have a mean 1 $\sigma_{rms}$ rms
noise level of $\sim$0.4 K per 0.06 km s$^{-1}$ channel on a T$_{mb}$
temperature scale that we use in this paper.  In the C$^{18}$O map,
the noise level and a slightly stripy appearance, resulting from the
on-the-fly mapping mode, becomes apparent. However, this does not
affect the PDFs because they are assembled from pixels above the
3$\sigma_{rms}$ level.  For further details, see Schneider et
al. (\cite{schneider2010}, \cite{schneider2011}).
\vspace{0.15cm}

\noindent {\bf Column-density maps from Herschel}\\  
\noindent  
For this study, we employ a column-density map at 36$''$ angular
resolution (regridded to 20$''$ to match the fully-sampled FCRAO maps)
that were obtained as part of the HOBYS ({\it Herschel} imaging survey
of OB Young Stellar objects) key program (Motte et al.
\cite{motte2010}).  This map was obtained by an SED (Spectral Energy
Distribution) fit to the PACS (Poglitsch et al.  \cite{poglitsch2010})
160 $\mu$m, and the SPIRE (Griffin et al.  \cite{griffin2010}) 250,
350, and 500 $\mu$m wavelengths observations. More details can be
found in Appendix A and a full map of the column density of the Cygnus
X North region is shown in Schneider et al. (\cite{schneider2015d})
and Bontemps et al.  (in prep.). A smaller cutout of the DR21 ridge at
25$''$ resolution is presented in Hennemann et
al. (\cite{hennemann2012}).

\section{The structure of the Cygnus X North region} \label{results}  

Cygnus X is one of the richest star-formation regions in the Galaxy,
containing the prominent OB-association Cyg OB2, with $\sim$50
O-stars (see Reipurth \& Schneider \cite{reipurth2008} for a review).
From large-scale $^{13}$CO 2$\to$1 (Schneider et al.
\cite{schneider2006}) and $^{13}$CO 1$\to$0 (Schneider et al.
\cite{schneider2007}) surveys, we derive a mass of a few 10$^6$
M$_\odot$ for the whole molecular cloud complex that is divided into
the Cygnus X North and South regions. These studies show that the
majority of the molecular clouds in the complex are located at a common distance of about
1.4--1.7 kpc. A distance of 1.4 kpc, which is derived from maser parallax (Rygl
et al.  \cite{rygl2012}), is now commonly accepted.
The clouds in the Cygnus X North region are actively forming
stars and contain more than 100 massive pre- and protostellar dense
cores (Motte et al.  \cite{motte2007}, Bontemps \cite{bontemps2010},
Csengeri et al. \cite{csengeri2011}, Hennemann et al. 2012).

Figure~\ref{herschel-co} shows the dust column-density map of Cygnus X
North with contours of $^{12}$CO 1$\to$0 overlaid onto it. The
$^{12}$CO emission is integrated over the velocity range -10 to 20 km
s$^{-1}$, which confines all star-forming molecular clouds that are
directly associated with the Cyg OB2 cluster. Outside this velocity
range, there are no other clouds along the line of sight, although in
other regions of Cygnus X, clouds from the Perseus arm appear at
velocities around -40 km s$^{-1}$.  As demonstrated in Schneider et
al. (\cite{schneider2006}, see their Fig.~3), there are mainly two
velocity coherent cloud complexes in Cygnus X North: the DR21
ridge-DR22-DR23 clouds at v=-10 to 1 km s$^{-1}$ and the W75N-DR17
clouds at v=7 to 20 km s$^{-1}$.  However, $^{12}$CO and dust also
trace more diffuse emission, mainly arising from the `Great Cygnus
Rift', a nearby (0.6-1 kpc) region of obscuration, which is identified
in optical images, and which is not associated with Cyg OB2.  As shown
in Schneider et al. (\cite{schneider2007}), the extinction due to the
Rift is of the order of \av $\sim$5--10 and mainly arises between v=6
to 20 km s$^{-1}$. Other authors determine values between \av=2--5
(Sale et al. \cite{sale2009}), \av=2.5--7 (Drew et
al. \cite{drew2008}), \av=5.5--7.5 (Wright et al. \cite{wright2010}),
and \av=2.6--5.6 (Guarcello et al. \cite{guarcello2012}). An
extinction around five is also the lowest emission level in the {\sl
  Herschel} dust map that is analysed in this paper. The Rift emission
is barely detected in $^{13}$CO and not in C$^{18}$O, CS, and
N$_2$H$^+$. This means that we expect only the low column-density
range of the $^{12}$CO PDF to be affected. The higher column-density
power-law tail is, in any case, not affected.

\begin{figure}[!htpb]  
\begin{centering}  
\vspace{-1.5cm}\includegraphics [width=11cm, angle={0}]{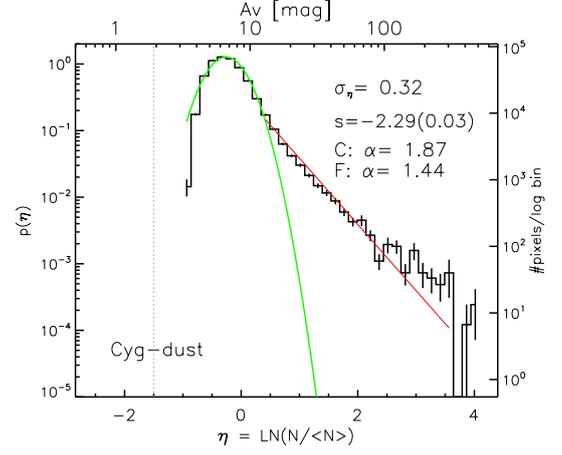}  
\vspace{-1.5cm}
\caption[] {PDF derived from the foreground-corrected {\sl Herschel}
  column-density map.  The left y-axis gives the normalized
  probability density $p(\eta)$, the right y-axis the number of pixels
  per log bin. The upper x-axis is the visual extinction and the lower
  x-axis the natural logarithm of the normalized column density.  The
  green curve indicates the best lognormal fit to the low column-density distribution, the red line displays a linear regression
  power-law fit to the high column-density tail. The dispersion of the
  fitted PDF ($\sigma_\eta$), the slope $s$ and its error, and the
  exponent $\alpha$ of an equivalent spherical density distribution
  (C) and a cylindrical density profile, such as for filaments (F), are
  indicated.}
\label{pdf-dust}  
\end{centering} 
\end{figure}  

\begin{figure}[!htpb]  
\begin{centering}  
\vspace{-1.5cm}\vspace{0cm}\includegraphics [width=9cm, angle={0}]{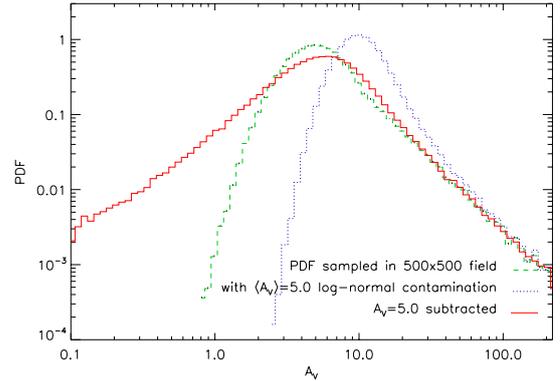}  
\caption[] {The original PDF of a model cloud (see text) with a
    lognormal core and a power-law tail is shown with a green dashed
    line. This cloud is `contaminated' with a model cloud that has a
    lognormal core. The resulting PDF is shown as a blue dashed
    line. The red line then displays the `corrected' PDF if a constant
    value of \av=5 is removed.}
\label{pdf-simu}  
\end{centering} 
\end{figure}

Figure~\ref{maps} shows the same cutout of Cygnus X North as
Fig.~\ref{herschel-co} in four different line tracers. The optical
depth of the lines decreases from left to right and the critical
density increases. While the CO 1$\to$0 lines have a low critical
density $n_{cr}$ of $\sim$10$^3$ cm$^{-3}$, the CS 2$\to$1 and the
N$_2$H$^+$ 1$\to$0 lines have $n_{cr}$ of 1.3$\times$10$^5$ cm$^{-3}$
and 6.1$\times$10$^4$ cm$^{-3}$, respectively, for a temperature of 10
K and thermal excitation (see. e.g. Flower \cite{flower1999}, Daniel
et al. \cite{daniel2005}, Shirley et al. \cite{shirley2015}). In
addition, CO depletes in cold and dense gas while CS, and in
particular N$_2$H$^+$, remain stable in this gas phase.  Accordingly,
the maps show selectively different density regimes of the gas. While
$^{13}$CO 1$\to$0 is still sensitive to lower-density gas and
resembles the $^{12}$CO map, N$_2$H$^+$ is confined to the densest
clumps.  Basically all clumps visible in the N$_2$H$^+$ map contain
cold, dense cores (starless and protostellar, indicated by crosses in
Fig.~\ref{maps}) observed in mm-dust continuum with the MAMBO
bolometer (Motte et al. \cite{motte2007}). From the submm-dust
observations, a typical size of 0.1 pc (0.7 pc) and density of 10$^5$
cm$^{-3}$ (10$^4$ cm$^{-3}$) for the cores (clumps) was derived.  This
value is confirmed by a decomposition of our N$_2$H$^+$ map using the
GAUSSCLUMPS algorithm (Stutzki \& G\"usten \cite{stutzki1990}) in the
velocity range -10 to 20 km s$^{-1}$, which shows an average clump
size of $\sim$0.6 pc. We only consider clumps larger than 1.5 times
the beamsize (45$''$).
   
\begin{figure}[!htpb]  
\begin{centering}  
\includegraphics [width=7.5cm, angle={0}]{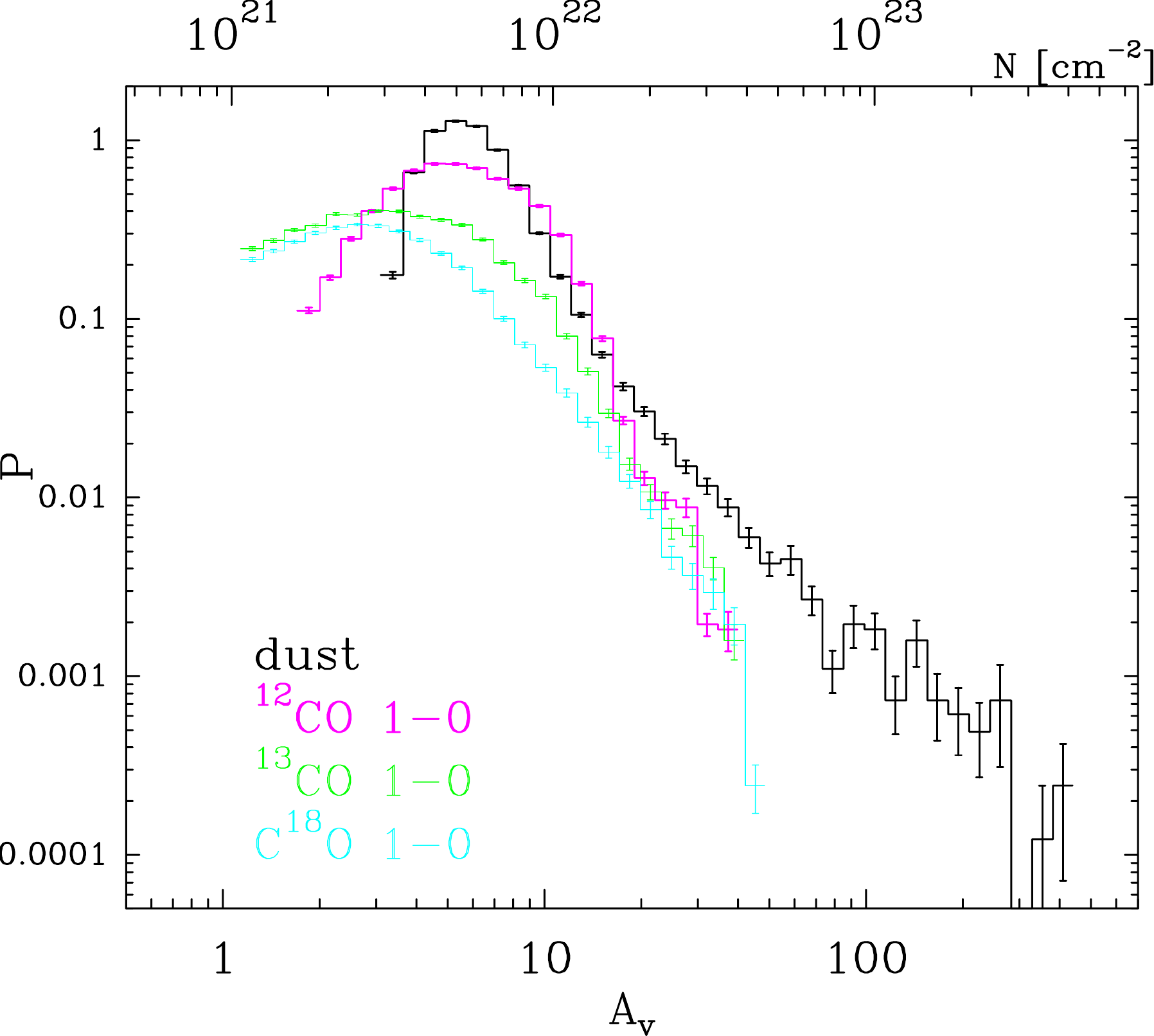}  
\caption[] {Probability distribution functions of H$_2$, obtained from
  dust and $^{12}$CO, $^{13}$CO, and C$^{18}$O 1$\to$0. All PDFs were
  constructed using pixels above the 3$\sigma$ noise level.}
\label{pdf-mol1}  
\end{centering}
\end{figure}

\begin{figure}[!htpb]  
\begin{centering}  
\includegraphics [width=7.5cm, angle={0}]{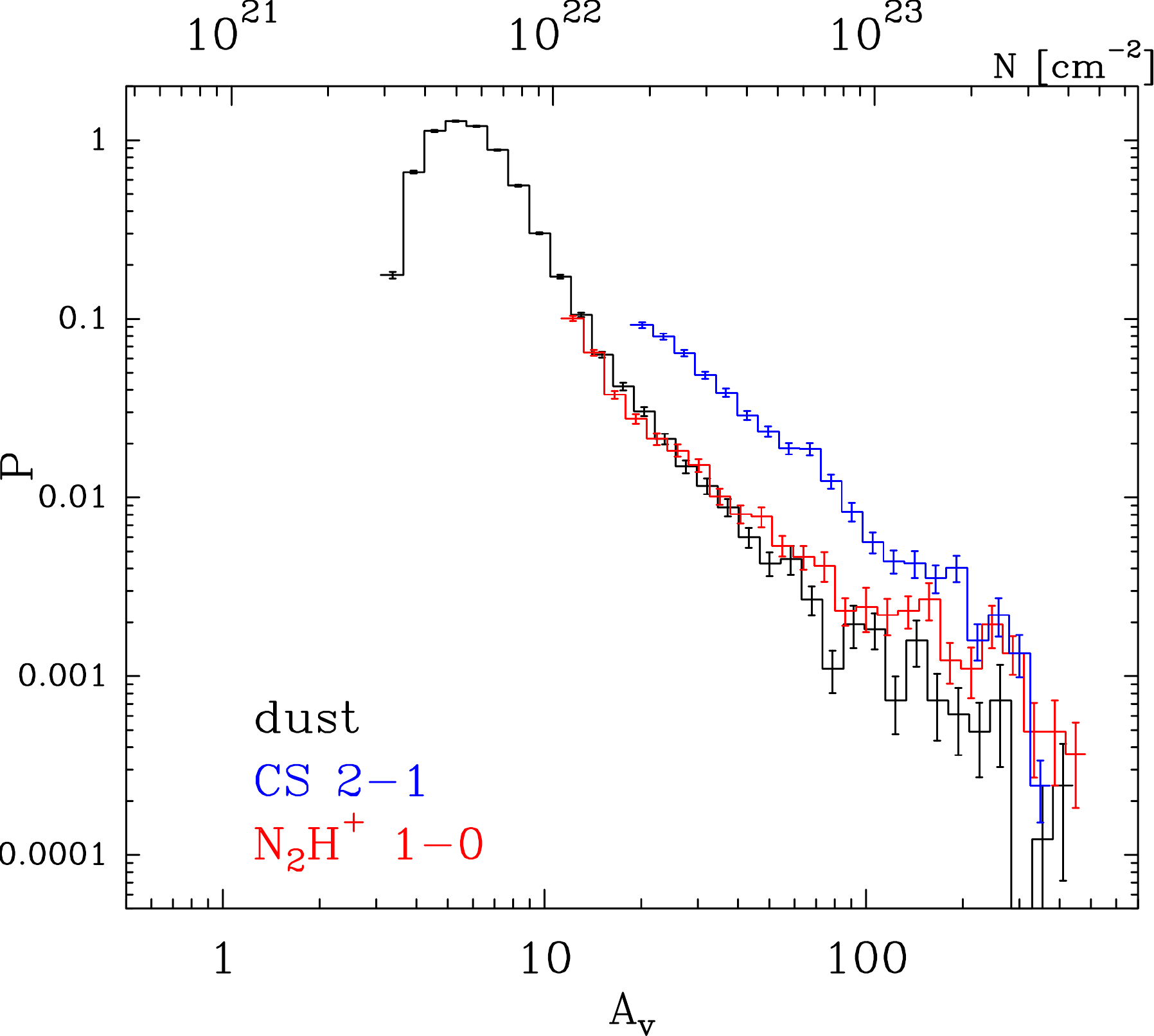}  
\caption[] {Like Fig.~4 but showing the CS 2$\to$1 and N$_2$H$^+$
  1$\to$0 PDFs (converted into H$_2$ column density) in addition to
  the dust PDF. }
\label{pdf-mol2}  
\end{centering} 
\end{figure}

\section{Probability distribution functions from dust and molecules} \label{extinction}   

\subsection{The dust PDF} \label{dust-PDF}

As discussed in the previous section, the dust map may suffer from
foreground contamination that mainly arises from the Cygnus Rift.
Contamination always modifies the measured cloud PDF. Relative to the
underlying PDF, the measured PDF becomes narrower, the peak shifts
towards higher \av, and the slope of the power-law tail steepens
(Schneider et al.  \cite{schneider2015a}).  Following the method
outlined in Schneider et al. (\cite{schneider2015a}), we can correct
for foreground contamination by removing the constant \av =5 value for
the Cygnus Rift from the original column-density map.

For a more realistic approach, we have to take into acount, however,
that the foreground contamination will typically also show a
distribution of column densities, usually also following a log-normal
distribution. In a systematic study, Ossenkopf et al.  (in prep.)
show that the correction by a constant foreground subtraction still
works quite well if the most probable contaminating column density is
used as the offset. This is illustrated in
Fig.~\ref{pdf-simu}. Following the description in Schneider et
al. (\cite{schneider2015a}), we contaminate a model cloud with the
parameters of the Cygnus PDF with a foreground cloud that is generated
from a fractional Brownian motion (fBm) map. This has a power spectral
index of 2.8, which matches the typically-observed spatial scaling
behaviour, and has a lognormal column-density PDF with a width
$\sigma_{\eta}=0.45$ and a contamination \av = 5.0.  
This cloud is thus a more realistic representation of a
foreground or background contaminating cloud. Figure~\ref{pdf-simu}
shows that the contamination produces a distortion in the PDF
but the simple correction obtained by subtracting a constant \av =5.0
value fully recovers the PDF tail and recovers the PDF peak position
approximately.

Even for such a relatively strong contamination, we can thus recover
the main PDF properties by subtracting a constant offset for the
contamination.  Figure~\ref{pdf-dust} shows the N-PDF obtained from
{\sl Herschel} dust continuum data, corrected for a contamination of
\av =5. This dust PDF shows the typical shape of PDFs obtained for
star-forming regions: a lognormal distribution for low column
densities followed by a power-law distribution between \av$\sim$10--15
and a few hundred \av .  Above \av $\sim$100, some excess in the PDF
is observed that could constitute a second power law, when compared to
a simple power law. Such an excess/flatter power-law tail was recently
found in some massive GMCs (Schneider et al. \cite{schneider2015c})
and was interpreted as a possible result of internal stellar
feedback. The same argument may hold here because only pixels in the
DR21 ridge, W75N, and some small, isolated clumps -- all associated
with young stellar objects (YSOs) -- contribute to this high
column-density part of the PDF (see Fig.~\ref{herschel-co}).

\begin{figure}[!htpb]  
\begin{centering}  
\includegraphics [width=7cm, angle={0}]{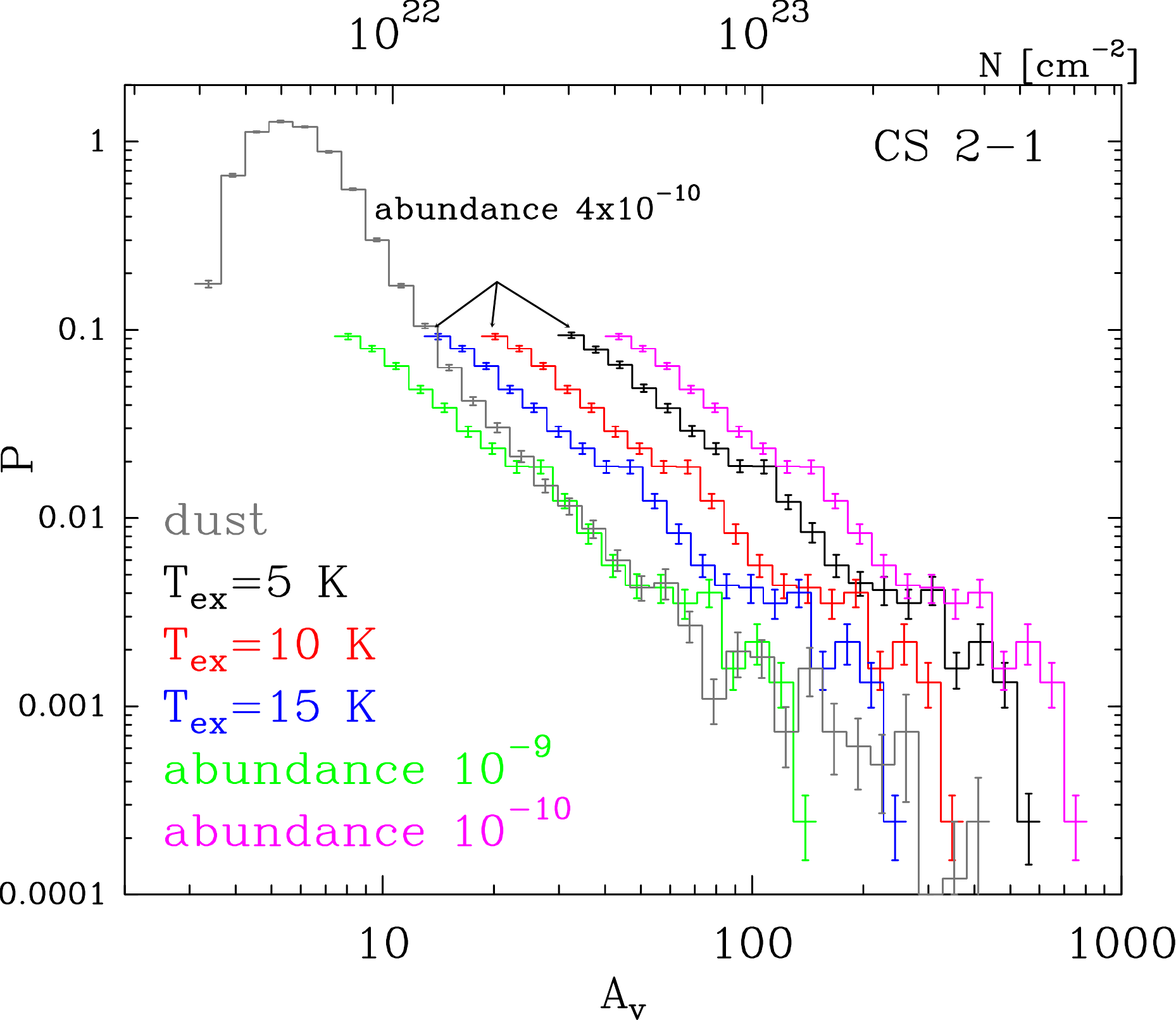}  
\caption[] {N-Probability distribution functions obtained from
  CS using (i) different excitation temperatures (5, 10, 15
  K) and an abundance of 4$\times$10$^{-10}$, and (ii) at a
  temperature of 10 K but abundance values of 10$^{-10}$ and
  10$^{-9}$. The dust PDF is shown for comparison.
}
\label{pdf-mol-tex}  
\end{centering} 
\end{figure}

\subsection{The PDFs from CO} \label{co-PDF}

The PDFs of H$_2$ column densities N$_{mol}$(H$_2$), which were
obtained from molecular lines, are shown in Fig.~\ref{pdf-mol1} and
\ref{pdf-mol2}, together with the same dust PDF from
Fig.~\ref{pdf-dust}.  Appendix B explains in detail how
N$_{mol}$(H$_2$) was derived.  In summary, we assume a constant
excitation temperature of 10 K, LTE conditions, and a
beam/source-filling factor of unity when determining all column
densities. We note, however, that PDFs are always resolution-limited.
The PDFs were constructed from pixels above the 3$\sigma$ noise level
and are normalized to the number of pixels from the dust map
(\#pix$_{dust}$), i.e. we scaled the molecular line PDFs by the factor
\#pix$_{mol}$/\#pix$_{dust}$. In Appendix C, we show the individual
PDFs without this scaling.  Above the noise level of \av$\sim$1--2,
the N$_{mol}$(H$_2$)-PDFs of $^{12}$CO, $^{13}$CO, and C$^{18}$O can
be fitted by a broad lognormal distribution with a peak at
\av$\sim$3--5, and by a power-law tail above \av=15 for
C$^{18}$O. Also, the $^{13}$CO PDF shows some excess above \av$\sim$15
(see Appendix C for a more detailed display).

Considering all uncertainties in the determination of H$_2$ column
density from dust and CO, the lognormal part in the CO-PDFs
corresponds quite well to the lognormal part of the dust PDF, which
indicates that the low-density molecular gas is probably well mixed
with dust. The $^{12}$CO PDF is a special case because it is
contaminated by line-of-sight confusion from the Cygnus Rift at a
distance of 600 pc, and thus shifts towards lower \av \ to some
degree.  For \av\ larger than $\sim$15 (we note that the column
densities are correct only within a factor $\sim$2), however the
$^{12}$CO PDFs depart from the lognormal shape and the distribution
becomes `bumpy' and falls off. This is the regime where the line no
longer traces the dense gas because it becomes optically thick.  In
comparison to hydrodynamic simulations with radiative transfer (Shetty
et al. \cite{shetty2011}), our observed $^{12}$CO PDF covers a lower
column-density range. Their `Milky Way cloud' (case (a) in Fig.~2)
$^{12}$CO PDF has the same width but peaks around \av=10-15.  More
recent results of Sz\"ucs et al. (\cite{laszlo2015}), however, show
$^{12}$CO PDFs that cover a very similar column-density range to the
one we observed.

\begin{figure*}[!htpb]  
\begin{centering}  
\includegraphics [width=8cm, angle={0}]{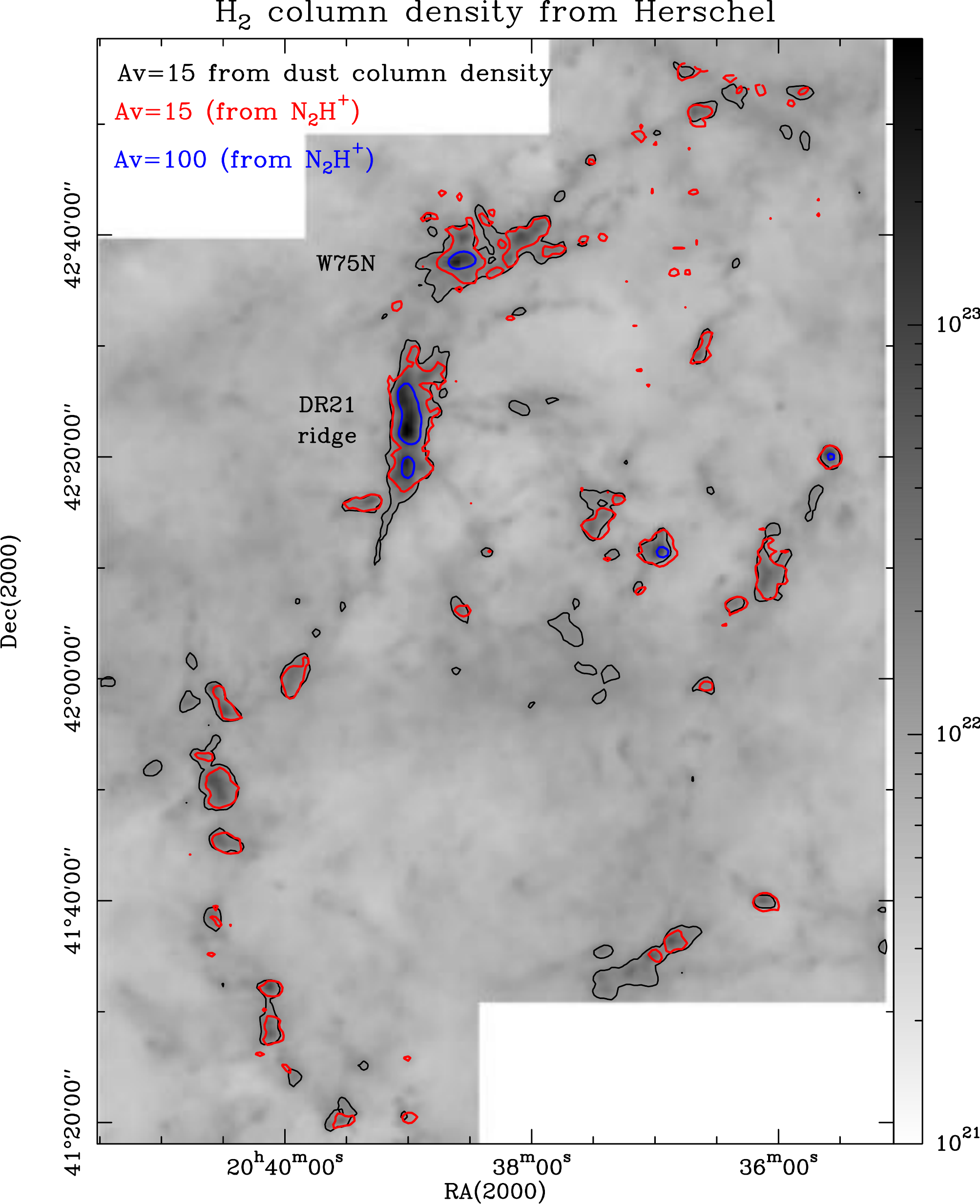}  
\includegraphics [width=7.9cm, angle={0}]{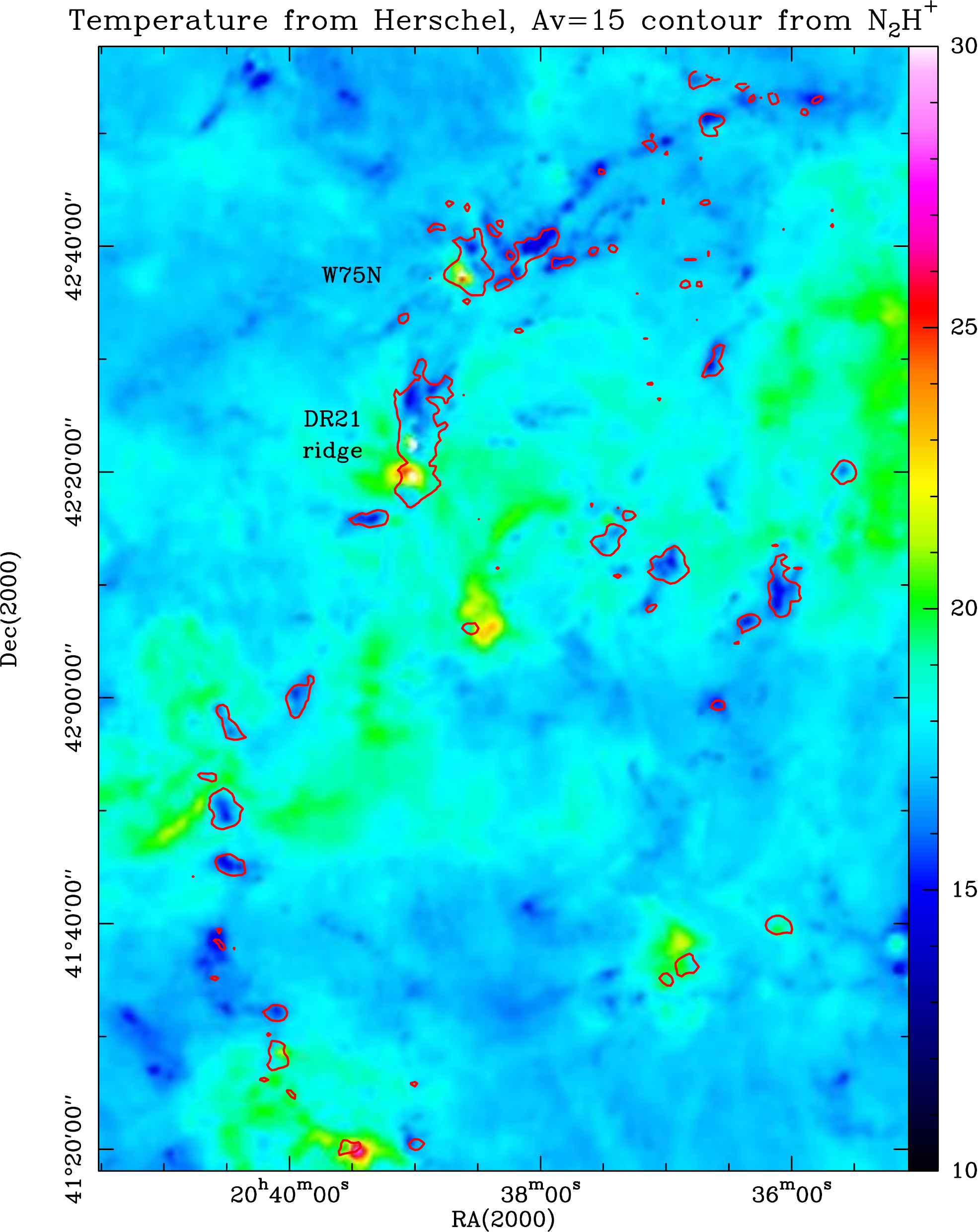}  
\caption[] {{\bf Left:} Foreground-corrected dust column-density map
  (as shown in Fig.~1) in greyscale.  The black contour indicates the
  \av = 15 level. The red contours show the same \av=15 level from the
  H$_2$ column-density map derived from N$_2$H$^+$, using an
  excitation temperature of 10 K and an abundance of
  2$\times$10$^{-9}$. {\bf Right:} Dust temperature map of Cygnus X
  North from {\sl Herschel} with the same red \av=15 contour from
  N$_2$H$^+$. }
\label{cyg-dust-n2h}  
\end{centering} 
\end{figure*}

The $^{13}$CO and C$^{18}$O PDFs extend to higher \av\ (up to
$\sim$40). The $^{13}$CO line is optically thin ($\tau\sim$0.1--0.3,
see Fig. B.3) in most of the clouds and becomes marginally more
optically thick ($\tau\sim$1) only in the DR21 ridge. The C$^{18}$O
line is optically thin everywhere with maximum values of $\tau\sim$0.4
in the DR21 ridge.  Optical depth effects, and not depletion, are the
main reason for the different PDF shapes of $^{12}$CO, $^{13}$CO, and
C$^{18}$O for higher column densities.  The power-law behaviour for
\av$>$15 that we find for C$^{18}$O indicates that the whole column of
gas is traced up to \av$\sim$40. Because $^{13}$CO becomes marginally
optically thin, we do not observe a clear power-law tail but only some
excess because the highest column densities are no longer
traced. $^{12}$CO becomes even more easily optically thick so that the
high column-density pixels are not traced at all and the PDF appears
lognormal.  Depletion of all CO isotopologues probably sets in for
\av\ larger than around 40 where all PDFs fall off.  The removal of CO
from the gas phase typically happens at densities above 10$^4$
cm$^{-3}$ at temperatures $\sim$15 K (e.g. Blake et
al. \cite{blake1995}). Our value of \av$\sim$40
($N$=3.8$\times$10$^{22}$ cm$^{-2}$) thus corresponds to a typical
clump size $d$ of $\sim$1.25 pc ($d = N/(10^4 {\rm cm}^{-3})$) in
which CO depletes. These clumps are the ones that are well traced by
N$_2$H$^+$ emission and the temperature map (Fig.~\ref{cyg-dust-n2h})
that we discuss in the next section.

At the low \av\ end, approximately in the range \av=1--2, we observe
some excess compared to a purely lognormal PDF distribution for
$^{13}$CO and C$^{18}$O (Fig.~C.1).  The chemistry in this
column-density regime is governed by (F)UV radiation via
photodissociation (\av$<$5) and fractionation (\av$<$3) processes
(e.g. van Dishoeck \& Black \cite{ewine1988}).  The excess indicates
that self-shielding of $^{13}$CO and C$^{18}$O against UV-radiation is
rather effective. Otherwise, the abundances would be lower and the PDF
would drop more significantly. On the other hand, in warmer regions
(internally or externally heated) in the medium \av\ range, CO
desorption leads to a higher abundance of CO.  These {\sl spatial}
abundance variations, which were observed for $^{13}$CO in Perseus
(Pineda et al. \cite{pineda2008}) and Orion (Ripple et
al. \cite{ripple2013}), have an obvious influence on the
N-PDF. However, they are difficult to assess (see next Section).
  
\subsection{The PDFs from CS and N$_2$H$^+$} \label{cs-PDF}

The N$_{mol}$(H$_2$)-PDFs of CS and N$_2$H$^+$ mostly cover the high
column-density range and can be best fitted by a pure power law (see
Appendix C). However, all absolute values need to be taken with
caution because they strongly depend on the abundance values that were
used for converting column densities into H$_2$ and on the excitation
temperature.

For N$_2$H$^+$, we adopt [N$_2$H$^+$]/[H$_2$] = 5$\times$10$^{-10}$, a
value determined for massive\footnote{Lower abundances around
  1-3$\times$10$^{-10}$ were found in low-mass dense cores and dark
  clouds (e.g. Caselli et al. \cite{caselli2002}, Bergin et
  al. \cite{bergin2001}, Tafalla et al. \cite{tafalla2006}).} cloud
cores (Pigorov et al. \cite{pigorov2003}), and applied it to one of
the Cygnus X North cores (N63, Motte et al. 2007) in the following way
(Fechtenbaum et al., in prep.): Level population diagrams that use
lines of the optically thin isotopologues $^{15}$N$_2$H$^+$ and
N$^{15}$NH$^+$ were plotted to estimate their column density. An
isotopic ratio $^{14}$N/$^{15}$N of 448 (Wilson \& Rood
\cite{wilson1994}) was used to derive the column density of
N$_2$H$^+$. The mass of 44 M$_\odot$ in a 30$''$ beam (Duarte-Cabral
et al. \cite{duarte2013}) of the source enabled us to calculate the
column density of H$_2$ and thus the abundance of N$_2$H$^+$.  The
resulting PDF covers a column-density regime of \av $\sim$15 to 400,
which corresponds very well to the one derived from dust (see
Fig.~\ref{pdf-mol2}).
  
For CS, there is a large scatter in observed values for high-mass
star-forming regions with [CS]/[H$_2$] = 10$^{-9}$ to a few times
10$^{-10}$ (Gianetti et al. \cite{gianetti2012}, Li et
al. \cite{li2015}, Neufeld et al. \cite{neufeld2015}). Low-mass cores
have typical values of a few times 10$^{-9}$ (e.g. Tafalla et
al. \cite{tafalla2006}).  Here, we use a value of 4$\times$10$^{-10}$
(Fechtenbaum et al., in prep.) that was obtained using the optically
thin isotopologue C$^{33}$S and applying the same method as outlined
above.  The PDF obtained with this abundance value covers the \av\,
range $\sim$20 to 350 and thus has a similar column density range as
the PDF determined from dust, but shifted vertically to the latter.
  
To investigate in more detail the effects of different excitation
temperatures and abundances, in Fig.~\ref{pdf-mol-tex} we plot a whole
set of PDFs for CS. We first vary the excitation temperature, using 5
and 15 K as two extreme cases, and 10 K (average value determined from
all molecular lines and dust). The column density variation is
significant, increases its value by more than a factor two between
T$_{ex}$=15 K and 5 K.  Different abundance values obviously have a
significant impact on the PDF as well.  A lower (but unlikely)
[CS]/[H$_2$] ratio of 10$^{-10}$ shifts the PDF to higher column
densities with unrealistic values of a few hundred \av\ .  A higher
abundance shifts the PDF towards lower column densities, and a value
of 10$^{-9}$ leads to a PDF that corresponds to the one obtained from
dust.  In this way, it is possible to `calibrate' the [CS]/[H$_2$]
abundance using the dust PDF. However, this method should be treated
with caution because the molecular line PDF also depends on the
excitation temperature, and the dust PDF has its own uncertainties
(variable temperature along the line of sight, unknown dust opacity
etc.).

\subsection{The high column density range in the maps  } \label{spatial}

The PDFs obtained from dust emission and N$_2$H$^+$ (and CS) both show
a power-law tail starting around
\av$\sim$15. Figure~\ref{cyg-dust-n2h} (left panel) illustrates the
high spatial correspondence between the pixels in the dust and
N$_2$H$^+$ maps that cover the same column-density range.  The black
and red contours indicate the \av\ = 15 level for the H$_2$ column
density derived from dust and N$_2$H$^+$, respectively. On average,
both contours cover the same areas, implying that the dust is well
mixed with the dense and cold gas that is traced by N$_2$H$^+$. The
geometry of the emitting regions is a mixture between spherical and
filamentary.  Even at very high column densities (indicated by the
arbitrary chosen \av=100 blue contour from N$_2$H$^+$), not only
spherical clumps are outlined but also the elongated DR21 ridge.  An
overlay between the dust temperature obtained with {\sl Herschel} and
the same \av=15 contour from N$_2$H$^+$ (Fig.~\ref{cyg-dust-n2h},
right panel) shows that it is indeed mostly cold gas traced by
N$_2$H$^+$ and dust. The clumps that are indicated by the red contour
typically have a temperature of 10-15 K, except those with internal
sources (the DR21 region in the DR21 ridge and DR23 at the southern
border of the map). In these cases, the column density is {\sl
  underestimated} so that the true power-law tails of both, the dust,
and N$_2$H$^+$ PDF, which is shown in Fig.~\ref{pdf-mol2}, are
slightly flatter.

\begin{table}[htbp] 
\begin{tabular}{lccccccc} 
\hline 
\hline 
Species            & $s$    & $\Delta s$  & $\alpha_c$ & $\Delta \alpha_c$ & $\alpha_f$ & $\Delta \alpha_f$ &    \\ 
                   &        &             &            &                   &            &                   & \\ 
\hline
Dust               & -2.29  & 0.03        & 1.87     &  0.03             &  1.44      &  0.03             & \\
C$^{18}$O 1$\to$0  & -2.57   & 0.12        & 1.78       &  0.04             &  1.39      &  0.04               & \\
CS 2$\to$1         & -1.56   & 0.02       & 2.28       &  0.02             &  1.64      &  0.02              & \\
N$_2$H$^+$ 1$\to$0 & -1.41   & 0.03        & 2.42       &  0.03            &   1.71      & 0.03              & \\
\end{tabular} 
\caption{Results from the power-law fit to the PDFs. $s$ and $\Delta
  s$ are the slope and its error, $\alpha_c$ and $\alpha_f$ are the
  exponents of a density distribution with $\rho \propto r^{-\alpha}$
  with spherical and filamentary geometry.}
\label{table} 
\end{table} 

\section{Discussion and conclusions} \label{discuss}

In the following section, we base our discussion on the column density
maps and PDFs of CS and N$_2$H$^+$ derived using an excitation
temperature of 10 K and abundance values of [CS]/[H$_2$] = 10$^{-9}$
(to be consistent with the dust PDF) and [N$_2$H$^+$]/[H$_2$] =
5$\times$10$^{-10}$, respectively.

\subsection{Thermal and subthermal excitation regimes}

For a homogeneous medium with beam-filling of unity, the H$_2$
column-density threshold $N$ for thermally excited N$_2$H$^+$ and CS
calculates from the clump size ($d$) and critical densities with $N =
d \, \times \, n_{cr}$. For $d$ we adopt 0.7 pc, taken from dust submm
observations (Motte et al. \cite{motte2007}).  This value is close to
the average clump size of 0.6 pc we derive from our N$_2$H$^+$ clump
decomposition\footnote{The submm continuum observations have an
  angular resolution of 11$''$, compared to the 45$''$ for N$_2$H$^+$,
  and trace a much larger column density regime. They are thus more
  precise for separating individual clumps.}. We thus obtain \av=285
and 130 for CS and N$_2$H$^+$, respectively, assuming critical
densities of 1.3$\times$10$^5$ cm$^{-3}$ (6.1$\times$10$^4$ cm$^{-3}$)
for CS (N$_2$H$^+$). Above these values, the emission lines should be
thermalized. However, the clumpiness of the gas leads to a lower
threshold of thermalization.  With an average density $\langle n
\rangle$=10$^4$ cm$^{-3}$ of the clumps,which was determined from
submm continuum observations (Motte et al. 2007), we obtain a beam
filling ($\langle n \rangle/n_{cr}$) of $\sim$10\% for CS and
$\sim$15\% for N$_2$H$^+$. Both lines are thus already thermalized at
lower \av\ .  For N$_2$H$^+$ we independently derive its thermal
excitation by the hyperfine-structure line ratios. The line is mostly
thermally excited because we obtain excitation temperatures above 5 K
for the emitting gas (Appendix C), which is above the typical value
for subthermal excitation.

CS has a higher critical density but depletes at densities above
$\sim$10$^5$ cm$^{-3}$, which is consistent with our PDF that is cut
off at \av$\sim$100 (for the abundance [CS]/[H$_2$] = 10$^{-9}$). On
the other hand, its chemistry is less density dependent. Hence, we
expect that CS is not thermalized at the lowest end of the
distribution.

\subsection{Slopes of the N-PDFs}

Independent of the exact column-density regime that is covered by the
PDFs from dust and CS/N$_2$H$^+$, they all show power-law tails with
slopes $s$ between --1.4 and --2.3 (see Table 1).  The slope $s$ of
the power law can be converted into the exponent $\alpha$ of an
equivalent density distribution $\rho \propto r^{-\alpha}$ (Appendix
D). For spherical geometry, representing a single core or a core
ensemble (Girichidis et al. \cite{giri2014}), we use
$\alpha_c$=1-2/$s$ (Federrath \& Klessen \cite{fed2013}).  For
singular polytropic cylinders, which portray filaments, we use
$\alpha_f$=1-1/$s$ (see Toci \& Galli (\cite{toci2015}) for regular
polytropic cylinder models and Fischera (\cite{fischera2014}) and
Myers (\cite{myers2015}) for the corresponding N-PDFs).  The values we
obtain for $\alpha$ ($\alpha_c$ or $\alpha_f$) vary between 1.4 and
2.4. In the simplified picture of the free-fall of a collapsing
sphere, $\alpha$=2 for early stages and $\alpha$ = 1.5 after a
singularity formed at the centre of the sphere (Shu \cite{shu1977},
Larson \cite{larson1969}, Penston \cite{penston1969}, Whitworth \&
Summers \cite{whitworth1985}). Our observed values from the dust
($\alpha_c$=1.9) and C$^{18}$O ($\alpha_c$=1.8) PDF slopes are
consistent with the spherical free-fall scenario, but CS and
N$_2$H$^+$ have higher values with $\alpha_c$=2.3 and $\alpha_c$=2.4,
respectively. All high-density clumps traced in CS and N$_2$H$^+$
emission are associated with pre- and protostellar sources (Motte et
al. 2007), which indicates that gravitational collapse in the
star-forming phase has started. The high column densities seen in
dust, CS, and N$_2$H$^+$ are thus not a consequence of a long column
of diffuse emission, but correspond to spatially concentrated high
volume densities.

On the other hand, the purely spherical free-fall picture seems
unlikely to apply to most of the gas which defines the N-PDF power
laws, because most of the gas we observe in Cygnus X is
filamentary. This is clearly illustrated in the maps shown in Figs.~1,
2, and 7, which display the filamentary network in the CO lines and,
with only a small fraction of gas in cores, that could be considered
roughly spherical (see CS and N$_2$H$^+$ maps in Figs.~2 and 7 and
submm continuum observations in Motte et al. 2007)\footnote{In Aquila
  (K\"onyves et al. \cite{vera2015}) the power-law tail contains more
  than $\sim$50\% of the mass in filaments but only $\sim$15\% in
  cores.}. Filaments can be in global free-fall collapse, as suggested
by molecular line observations (e.g. Peretto et
al. \cite{peretto2013}, Schneider et al. \cite{schneider2010},
\cite{schneider2015b}) but the gas we see in filaments (and cores) can
account for power-law N-PDFs if it has the right power-law
radial-density structure, even in hydrostatic balance with no net
inward motion. The observed N-PDF slopes ($s$=-1.4 to -2.6, leading to
$\alpha_f$ between 1.4 and 1.7) are consistent with self-gravitating
but non-collapsing filament models (Myers \cite{myers2015}, Toci \&
Galli \cite{toci2015}). In this case, filamentary gas that has a
sufficient mass per length ratio is centrally concentrated and
self-gravitating, but does not need to be globally collapsing.
However, to confine the filaments/clumps, an external pressure
$p_{ext} \propto \langle n \rangle \, T_{kin}$ of the order of
$p_{ext} \sim 10^{4-5} \, {\rm cm^{-3}} \times 10 \, {\rm K}$ is
required (where $\langle n \rangle$ is the average density in the
filaments/clumps and $T_{kin}$ the gas temperature).  However, this
pressure cannot be isotropic, provided by the lower density envelope
that is well traced in $^{12}$CO at a density of $\sim$10$^3$
cm$^{-3}$ and a temperature of $\sim$10--40 K. Moreover, it could be
ram pressure of new gas that is accreting onto filaments and clumps, a
process that is now frequently observed (e.g. Schneider et
al. \cite{schneider2010}, Kirk et al. \cite{kirk2013}, Palmeirim et
al. \cite{pedro2013}).

The flatter power-law slopes of the CS and N$_2$H$^+$ PDF in
comparison to the dust could reflect an abundance change because
N$_2$H$^+$ is produced when CO is depleted at very high column
densities (Bergin et al. \cite{bergin2001}, Tafalla et al. 2002,
2006).  In addition, because CS is not thermalized at the lowest
column densities, this could lead to a somewhat flatter tail with
respect to the dust.

In summary, a power-law distribution in the density structure is
required to have a power-law tail in the PDF and this can be
achieved by either a hydrostatic configuration, where the power law
arises from a balance of gravitational forces and pressure gradients,
or directly in a dynamically collapsing system.  Probably both play a
role, with filaments possibly balanced or at least contraction-slowed
by (magnetic) pressure gradients on the one hand and very dense collapsing
cores on the other.

\section{Summary}

We derived probability distribution functions (N-PDFs) of H$_2$ column
density for the Cygnus X North region from dust, $^{12}$CO, $^{13}$CO,
C$^{18}$O 1$\to$0, CS 2$\to$1, and N$_2$H$^+$ 1$\to$0. The
determination of the H$_2$ column density from the molecular lines is
based on standard procedures and abundances, assuming a common
excitation temperature of 10 K and LTE. Our findings are:

\noindent $\bullet$ The PDFs of dust and CO isotopologues can
be described by a lognormal distribution for the low column-density range
between \av$\sim$1 to $\sim$15, with a common peak around \av$\sim$5. 
Though line-of-sight contamination and variations in abundance and excitation
temperatures introduce large uncertainties in the CO-PDFs, the overall
correspondence between these observational PDFs and the ones from recent
simulations (Sz\"ucs et al. 2015) is good. \\
\noindent $\bullet$ Optical depth effects are the main reason for the
different PDF shapes of $^{12}$CO, $^{13}$CO, and C$^{18}$O for higher
column densities.  The PDF from $^{12}$CO is cut above \av$\sim$30
because the line becomes optically thick.  The marginally optically
thin $^{13}$CO line shows excess in its PDF for \av=15-40, while the
PDF for the optically thin C$^{18}$O line displays a power-law tail in
the same \av=15-40 range. \\
\noindent $\bullet$ Neither selective photodissociation nor 
fractionation seem to play a significant role for the CO abundances
for low column densities. \\
\noindent $\bullet$ Depletion of all CO isotopologues probably sets in
for \av\ greater than around 40 where all PDFs fall off. \\
\noindent $\bullet$ The PDFs of CS and N$_2$H$^+$ only consist of a
power-law tail that covers a high column density range (\av$\sim$15 to
a few 100 \av). Using CS as an example, we discuss the potential influence of
  abundance and excitation temperature variations, and show that these
  could shift the entire PDF by more than one magnitude in column
  density. To better constrain this uncertainty, we `calibrate' the
  molecular line PDF using the dust-emission map (i.e. shifting the
  distribution at a given excitation temperature so that it
  corresponds to the dust PDF). For CS, we obtain an
  abundance of [CS]/[H$_2$]=10$^{-9}$ in this way. \\
\noindent $\bullet$ We find that the N$_2$H$^+$ and CS lines are mainly
thermally excited. N$_2$H$^+$ is well mixed with the dust and traces spatially
the same dense gas clumps of a typical size of $\sim$0.6-0.7 pc that were shown to contain
pre- and protostellar sources (Motte et al. 2007). \\
\noindent $\bullet$ The slopes of the high column-density power-law tail end
of the N-PDFs from dust, CS, and N$_2$H$^+$ are $s$=-2.3, -1.6, and -1.4, respectively. 
These values are consistent with a gas distribution that is dominated by gravity, i.e.
free-falling gas in cores and filaments, although a hydrostatic configuration with ram
pressure by gas accretion can also take place. The  power law then arises
from a balance of gravitational forces and pressure gradients.   

Summarising the observational results, we find that the N-PDFs
obtained from molecular lines are not well confined because they
depend strongly on excitation temperature and abundance, and various
combinations of these can lead to the same PDF. `Calibrating' a
molecular line PDF with dust is an appealing approach but it should be
kept in mind that dust PDFs also suffer from uncertainties. For
example, the specific dust opacity is not constant (a value of
$\beta$=2 was chosen for this paper) and line-of-sight contamination
can lead to an overestimation of the column density.  However, this
study nevertheless shows that the dust provides the highest dynamic
range in tracing the low to high column-density regime, compared to
the molecular line tracers (see also Goodman et al. \cite{good2009}
and Burkhart et al. \cite{burk2013}).

\begin{acknowledgements}  
  N.S. and S.B. acknowledge support by the ANR-11-BS56-010 project
  STARFICH. N.S., V.O., and T.C. acknowledge support from the Deutsche
  Forschungsgemeinschaft, DFG, through project number 0s 177/2-1 and
  177/2-2, and central funds of the DFG-priority program 1573
  (ISM-SPP). C.F. acknowledges funding provided by the Australian
  Research Council’s Discovery Projects (grants DP130102078 and
  DP150104329). R.S.K. acknowledges subsidies from the DFG priority
  program 1573 (Physics of the Interstellar Medium) and the
  collaborative research project SFB 881 (The Milky Way System,
  subprojects B1, B2, and B5).
  \end{acknowledgements}

\newpage
  
\appendix     

\section{Determination of H$_2$  column density from dust}

We assume optically thin dust emission at one dust temperature $T_d$
for each pixel in the map and fitted a greybody function of the form
$I_\nu=B_v(T_d)\,\kappa\,\Sigma$ to the absolutely calibrated
datapoints at 160 $\mu$m, 250 $\mu$m, 350 $\mu$m, and 500 $\mu$m.  The
specific dust opacity per unit mass (dust+gas) is approximated by the
power law $\kappa_\nu \, = \,0.1 \, (\nu/1000 {\rm GHz})^\beta$
cm$^{2}$/g with $\beta$=2, and the dust temperature and surface
density distribution $\Sigma$ are left as free parameters. The H$_2$
column density $N_{dust}$ is then calculated from
$\Sigma=\mu_H\,m_H\,N_{dust}$, adopting a mean molecular weight per
hydrogen molecule $\mu_H$=2.8. For more details, see Hill et
al. (2011).  We approximate the final uncertainties in the dust
column-density maps to be around $\sim$30--50\%, mainly as a result of
the uncertainty in the assumed form of the opacity law and possible
temperature gradients along the line of sight.

\section{Determination of H$_2$ column density from molecular lines}

\begin{figure}[!htpb]  
\begin{centering}  
\includegraphics [width=8cm, angle={0}]{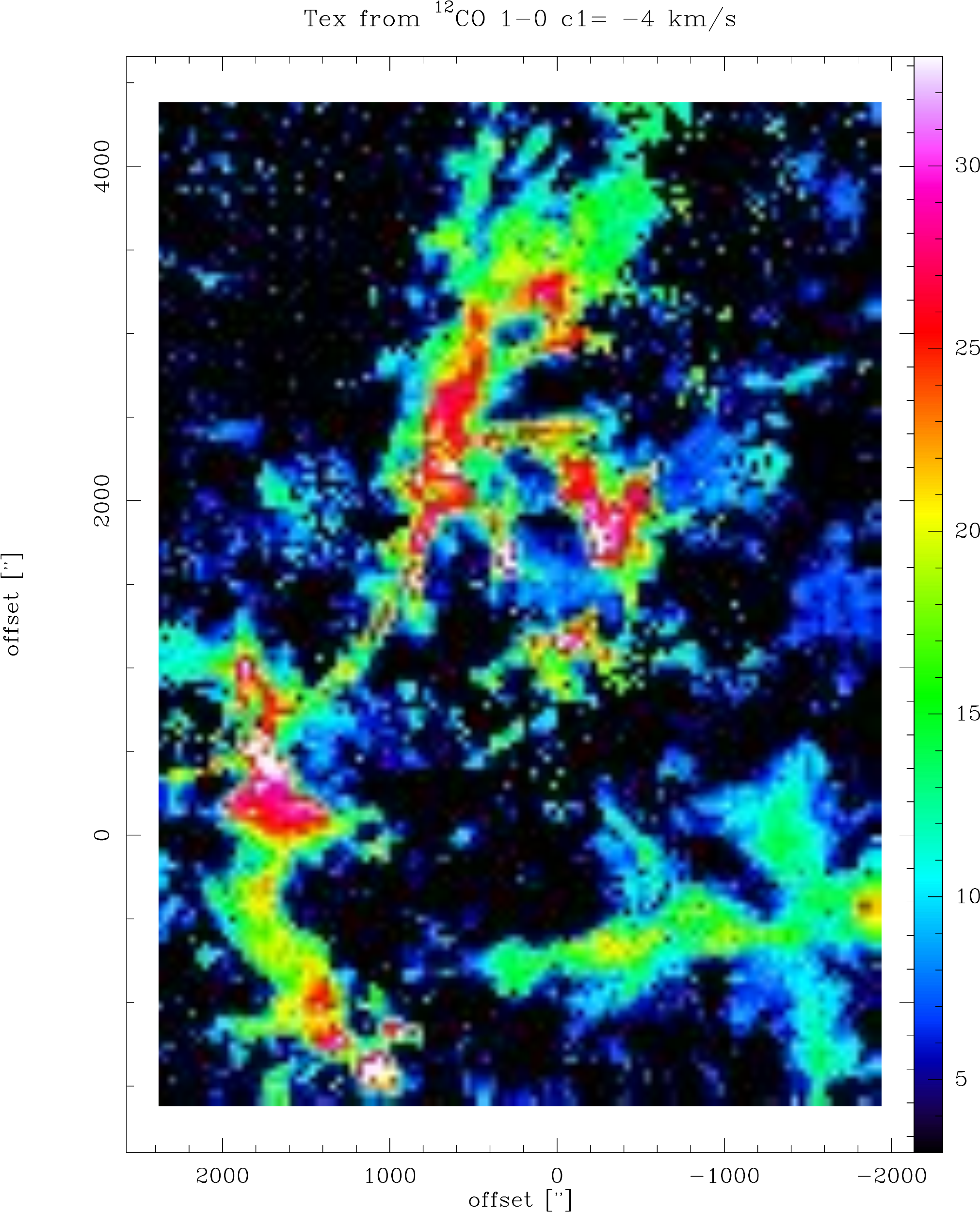}  
\caption[] {Map of excitation temperature of the main velocity component in
  Cygnus X North obtained from a Gaussian fit to the $^{12}$CO 1$\to$0 line.} 
\label{map-tex}  
\end{centering} 
\end{figure}

\subsection{Determination of and constraints on the excitation temperature} 

The temperature stucture in the Cygnus X North region is complex.
Large-scale UV illumination from the central Cyg OB2 cluster heats the
diffuse gas and PDR surfaces.  Internal (proto)-stars and clusters
such as DR21 show up as local hotspots. In parallel, the Cygnus X
North clouds also contain a large reservoir of cold and dense gas,
seen in submm continuum (Motte et al. \cite{motte2007}).

The observations of several molecular lines enable us to trace
sensitively this temperature structure. If the LTE assumption is valid
and the density is larger than the critical density of the transition,
the kinetic temperature of the gas (T$_{kin}$) corresponds to the
excitation temperature T$_{ex}$ for the molecular lines (which is
assumed to be equal for all species). If gas and dust are well mixed,
this temperature should also correspond to the dust temperature
T$_{dust}$ that was derived from {\sl Herschel}.  However, LTE
conditions are not present and the density varies from low values in
the diffuse gas phase, mainly seen in $^{12}$CO, to high values in
dense clumps in the molecular clouds, traced by dust and the CS and
N$_2$H$^+$ lines.  In the following, we determine the excitation
temperature in different ways, using the observational data sets. \\

\noindent {\bf T$_{ex}$ from $^{12}$CO 1$\to$0} \\
From $^{12}$CO 1$\to$0, we calculate the excitation temperature assuming
that this line is optically thick so that 
\begin{eqnarray} 
{\rm T}_{ex}(^{12}{\rm CO}) & = & \frac{5.53}{\ln(1+5.53/({\rm T}_{mb}(^{12}{\rm CO})+0.818))}  
.\end{eqnarray}
The molecular clouds in the Cygnus region consist of several velocity
components (Schneider et al.  \cite{schneider2006}). We performed a
Gaussian line fit to the  four main lines (see explanations for the
$^{13}$CO column density). Figure~\ref{map-tex} shows the resulting
excitation temperature for the most prominent component at velocities
around -4 km s$^{-1}$.  It becomes obvious that the temperature
varies significantly between $\sim$5 and $\sim$40 K. \\

\noindent {\bf T$_{ex}$ from N$_2$H$^+$ 1$\to$0} \\
We calculate the excitation temperature from N$_2$H$^+$ by fitting its
hyperfine structure with the known relative intensities of the seven
components. The fit delivers
simultaneously the opacity $\tau$ and T$_{ex}$ and shows no anomaly
in the relative intensities.  The average value of T$_{ex}$ is 7 K (in
a range of 5 to 20 K) and the opacity is generally below one. However, we emphasise
that only the product $\tau \times$T$_{ex}$ is constrained by the fit
so that the excitation temperature is not well constrained. \\ 

\noindent {\bf T$_{ex}$ from dust} \\
The right-hand panel of Fig.~6 shows the dust temperature we derive from
the SED fit to the {\sl Herschel} fluxes.  The average value for the
dust temperature across the map is 15 K, with a variation between 10
and 25 K (see also Hennemann et al. 2012). Because the dust
temperature is an average along the line of sight, it is not well
determined in regions with strong temperature gradients. However, for
cold, dense clumps and cores that are associated with CS and N$_2$H$^+$
emission, we expect a much smaller variation in temperature. \\

In summary, considering the excitation temperatures determined from
the molecular lines and the dust, we use 10 K in the following as the
best compromise for the determination of column densities. However, we
also discuss other methods (using T$_{ex}$ from $^{12}$CO) to
obtain the $^{13}$CO column density, and study the influence of
different T$_{ex}$ for the N$_2$H$^+$ column density.

\subsection{Column densities from linear molecules} 

The beam-averaged total column density $N$ of any optically thin molecule can be
determined from the observed line-integrated main-beam brightness
temperature $T_{mb}$ with
\begin{eqnarray} 
{\rm N} \, [cm^{-2}] & = & f(T_{ex}) \int {\rm T}_{mb} [K] \, d{\rm v} [km s^{-1}] 
\nonumber \\  
\end{eqnarray} 
and for linear molecules    
\begin{eqnarray} 
f(T_{ex}) & = & \frac{3hZ}{8 \pi^3 \mu^2 J_u} 
\frac{\exp(E_{up}/kT_{ex})}{[1-\exp(-h\nu/kT_{ex})] (J(T_{ex}) - J(T_{BG}))}  
\nonumber \\   
& &    
\label{ncol}   
\end{eqnarray} 
and 
\begin{eqnarray} 
J(T_{ex}) & = & \frac{h\nu}{k(\exp(h\nu/(kT_{ex})-1)}  
\end{eqnarray} 
and $J(T_{BG}) = J(2.7K),$ and 
in which $h$ and $k$ denote the Planck and the Boltzman constants,
respectively, $E_{up}$ is the energy of the upper level, $\nu$ is the
frequency [GHz], $\mu$ is the dipole moment [Debye], $J_u$ is the
upper value of the rotational quantum number and $\int {\rm T}_{mb}\,
d{\rm v}$ is the velocity integrated line intensity on a main beam
temperature scale. 

The first two terms of the rotational partition function for a diatomic linear molecule which is
accurate to 1\%, compared to the full term for temperatures above 2 K (Mangum \& Shirley 2015), are given by  
\begin{eqnarray} 
Z & = & \frac{kT_{ex}}{h B}+1/3  
,\end{eqnarray} 
with the rotational constant B expressed to first order as $\nu$=2BJ$_u$. 

To determine all H$_2$ column densities, we apply a
correction to the hydrogen mass of a factor of 1.36 to account for
helium and other heavy elements. \\

\begin{figure*}[!htpb]  
\begin{centering}  
\vspace{0cm}\includegraphics [width=7cm, angle={0}]{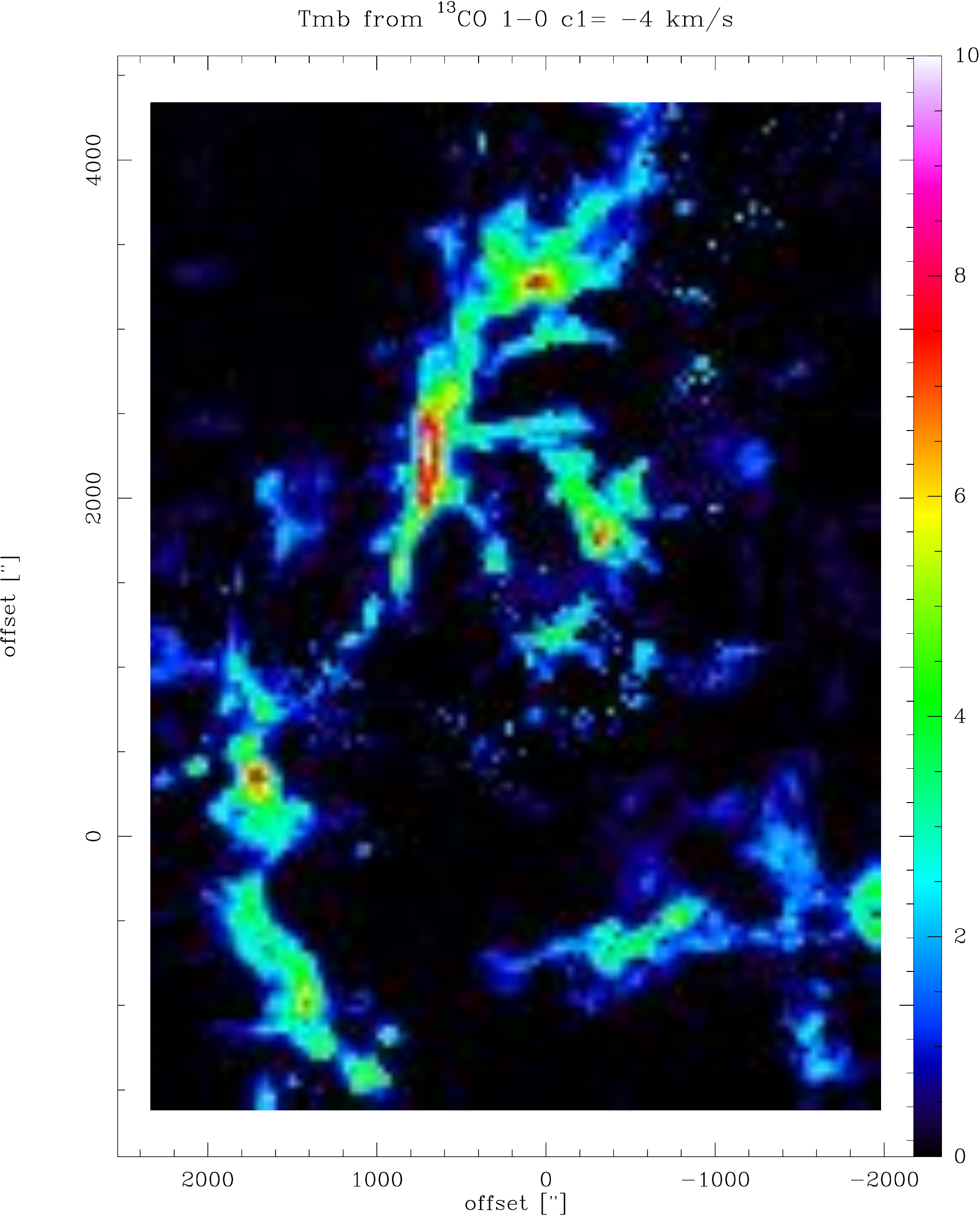}  
\vspace{0cm}\includegraphics [width=7cm, angle={0}]{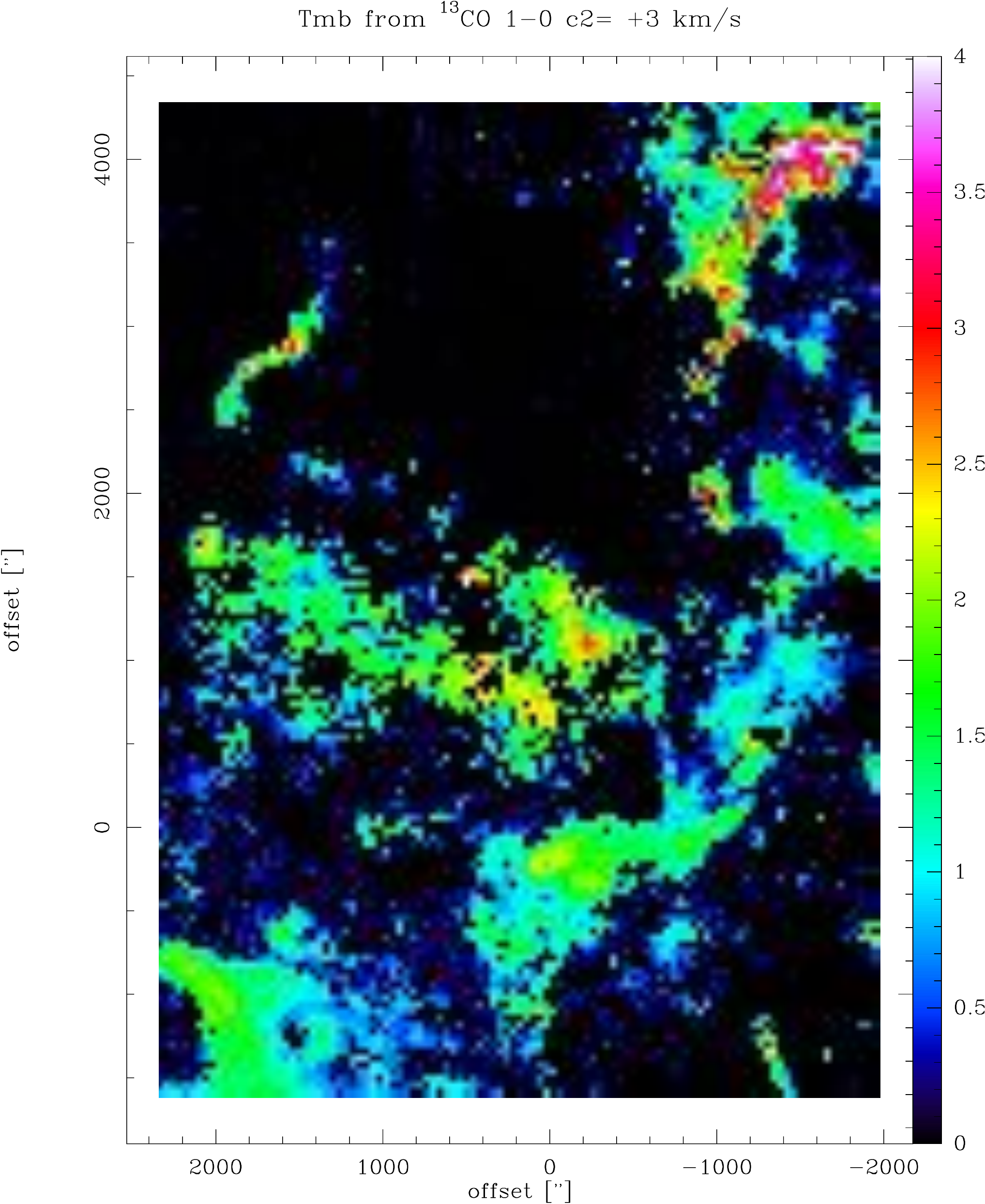}  
\vspace{0cm}\includegraphics [width=7cm, angle={0}]{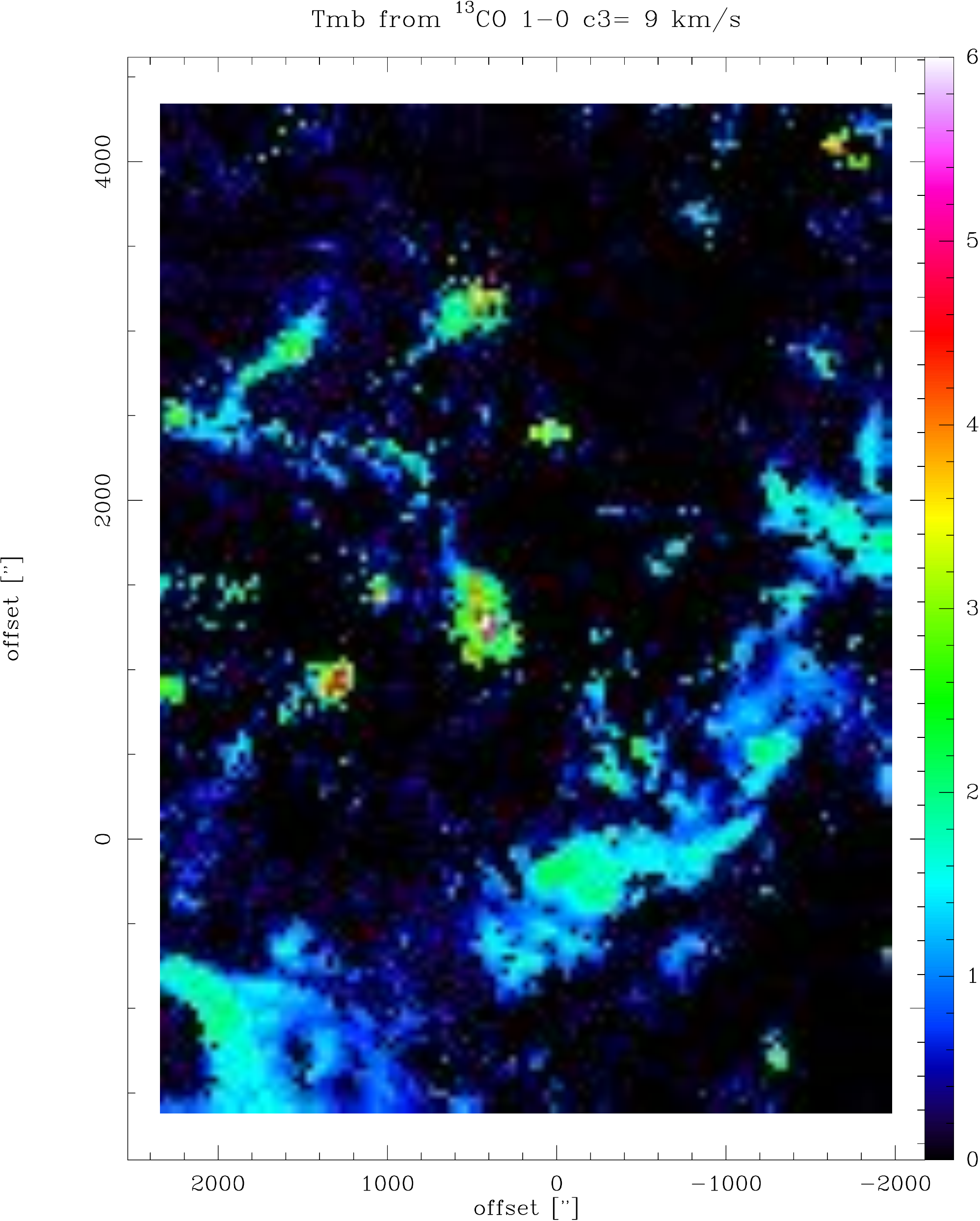}  
\vspace{0cm}\includegraphics [width=7cm, angle={0}]{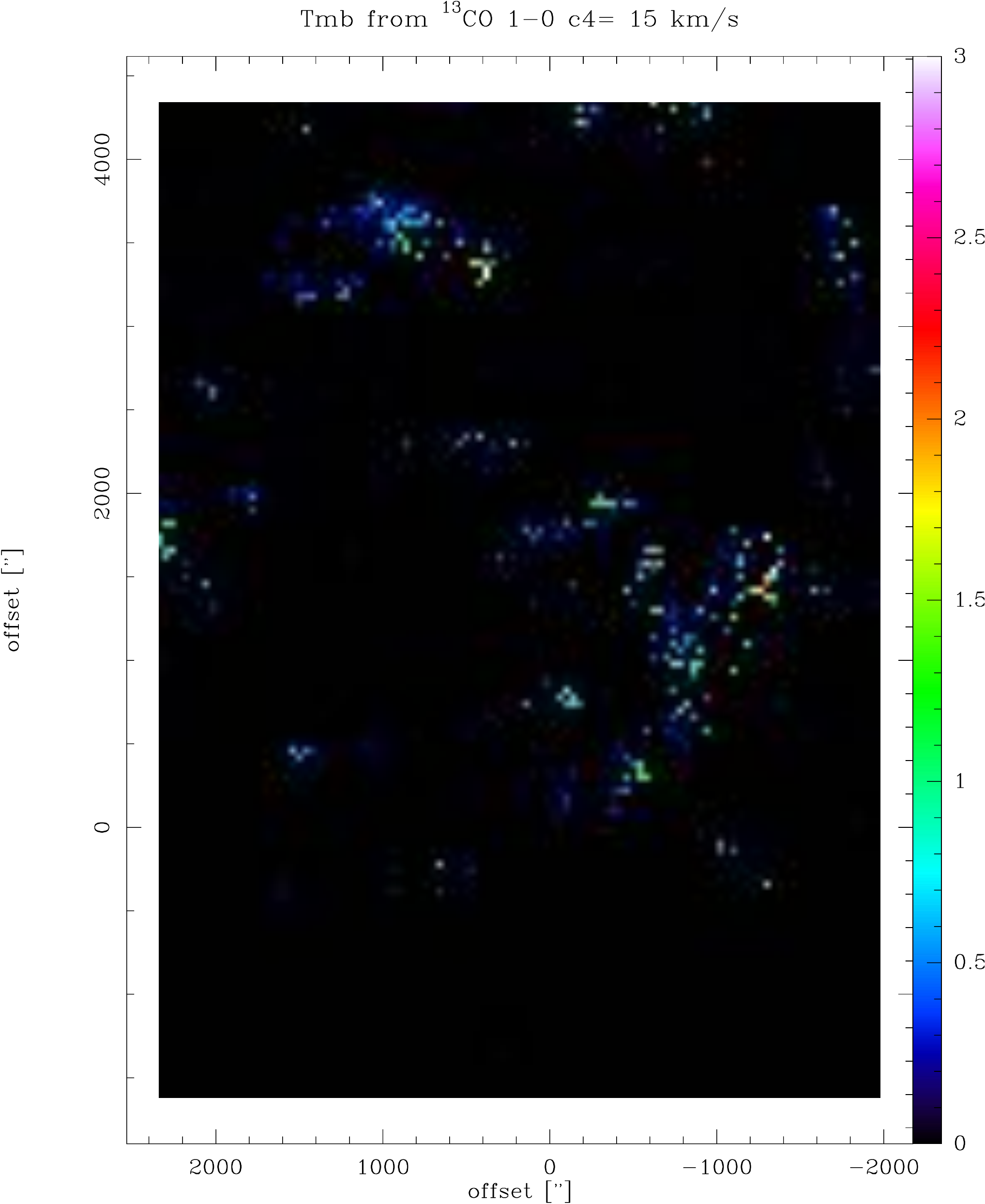}  
\caption[] {Maps of the Gaussian fits to the main-beam brightness temperature
  of $^{13}$CO 1$\to$0 for the four main velocity ranges in Cygnus X North.
  The most prominent emission comes from the velocity component at -4 km s$^{-1}$.}  
\label{maps-13co}  
\end{centering} 
\end{figure*}  

\begin{figure}[!htpb]  
\begin{centering}  
\vspace{0cm}\includegraphics [width=6cm, angle={0}]{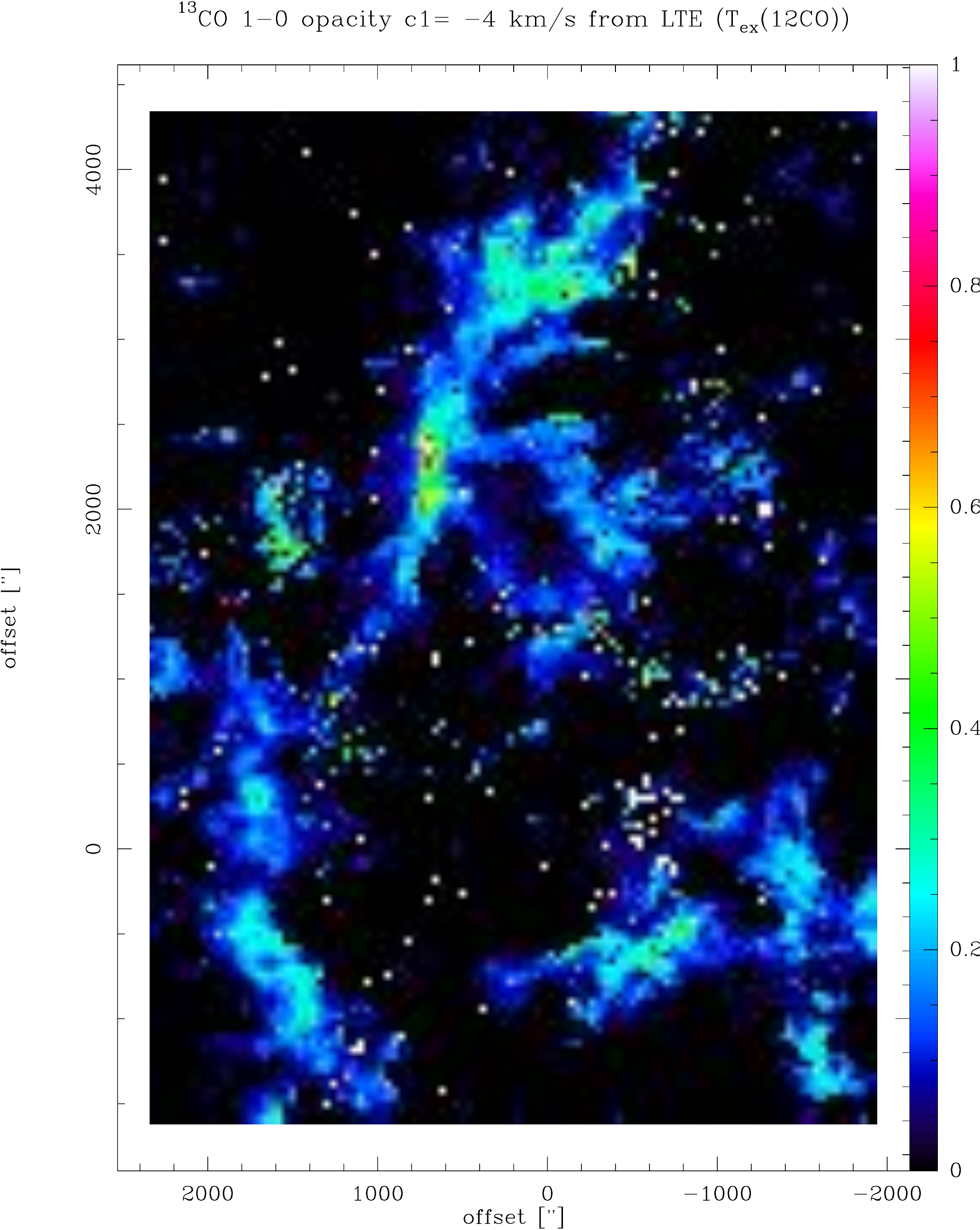}  
\caption[] {Map of the $^{13}$CO opacity derived for the velocity
  component -4 km s$^{-1}$. White pixels indicate positions that were
  not well-fitted (e.g. FWHM too large, intensities too low) and thus
  attributed a blanking value.}
\label{map-13co-tau}  
\end{centering} 
\end{figure}  

\noindent{\bf {H$_2$ column density from $^{13}$CO 1-0}} \\ 
\noindent The values for the $^{13}$CO 1$\to$0 transition are
$\nu$=110.201 GHz, h$\nu$/k=5.29 K, $\mu$=0.112 Debye, $J_u$=1, and
$J(2.7K)$=0.868. The temperature-dependent factor f(T$_{ex}$) is
3.22, 2.00, 1.62, 1.58, and 1.70$\times$10$^{15}$ K$^{-1}$
(km/s)$^{-1}$ cm$^{-2}$ for 5, 7, 10, 15, and 20 K, respectively.
As discussed above, we use a constant value of 10 K for T$_{ex}$.
A change in the excitation temperature between 5 and 15 K implies only
a 10\% variation in the column-density value.  We then use an
abundance [$^{12}$C]/[$^{13}$C]=70, determined from Wilson \& Rood
(\cite{wilson1994}). This value corresponds well to the one derived for
Orion A (Langer \& Penzias \cite{langer1990}). For the
[H$_2$]/[$^{12}$CO] abundance, we use a value of 1.1 10$^4$
(Pineda et al. \cite{pineda2010}, Fontani et al. \cite{fontani2012}).

Because it is known that the $^{13}$CO 1$\to$0 line is not always
optically thin, an opacity correction $\tau_{13}/(1-\exp{(\tau_{13})}$)
for moderate $\tau$ can be applied for the column density (Frerking
et al. \cite{freking1982}). To verify if this kind of correction is
required, we determine the $^{13}$CO opacity by using the excitation
temperature T$_{ex}$ that was obtained from $^{12}$CO and the observed
$^{13}$CO main-beam brightness temperature T$_{mb}$($^{13}$CO) with
\begin{eqnarray}
  \tau_{13}  =  -\ln(1-{\rm T}_{mb}(^{13}{\rm CO})/(5.29/(\exp(5.29/{\rm T}_{ex})-1)-0.868)). \nonumber \\  
\end{eqnarray}
Because the molecular clouds in the Cygnus region consist of several
velocity features (Schneider et al.  \cite{schneider2006}), we
perform a Gaussian line fit to four components for $^{12}$CO and
$^{13}$CO (see B.1). As shown in Fig.~\ref{maps-13co}, the dominating
emission comes from the component at -4 km s$^{-1}$, which is
associated with the molecular clouds of DR21, DR22, and DR23. The
component at 9 km s$^{-1}$ becomes important for the W75 region and
southwest of the DR21 ridge.  Emission at 3 and 15 km s$^{-1}$ is more
diffuse and on a low intensity level.  We calculate the $^{13}$CO
opacity for all velocity components and find that the line is
optically thin everywhere, even for the -4 km s$^{-1}$ component that
is shown in Fig.~\ref{map-13co-tau}. We thus did not include an
opacity correction for the determination of the $^{13}$CO column
density.

The N-PDFs that were determined from the $^{13}$CO column density,
using a constant excitation temperature of 10 K or a variable
pixel-to-pixel temperature T$_{ex}$ from $^{12}$CO, do not show a
significant difference. \\

\noindent{\bf {H$_2$ column density from C$^{18}$O 1-0}} \\ 
\noindent The values for the C$^{18}$O 1$\to$0 transition are
$\nu$=109.782 GHz, h$\nu$/k=5.27 K, $\mu$=0.1098 Debye, $J_u$=1, and
$J(2.7K)$=0.872 K. The temperature-dependent factor f(T$_{ex}$) is
3.34, 2.08, 1.69, 1.65, and 1.78$\times$10$^{15}$ K$^{-1}$
(km/s)$^{-1}$ cm$^{-2}$ for 5, 7, 10, 15, and 20 K, respectively. As discussed
in Sec. B.1, we adopt a value of 10 K for T$_{ex}$.  To determine the
H$_2$ column density, we then use a ratio [H$_2$]/[$^{12}$CO] of 1.1
10$^4$ (Pineda et al. \cite{pineda2010}, Fontani et
al. \cite{fontani2012}) and an abundance [$^{16}$O]/[$^{18}$O] = 531,
which was determined from Wilson \& Rood (\cite{wilson1994}), using a distance of
1.4 kpc.

We determine the C$^{18}$O opacity in the same way for the four line
components as we did for $^{13}$CO (see above), using the excitation
temperature T$_{ex}$ obtained from $^{12}$CO and the observed
C$^{18}$O main-beam brightness temperature T$_{mb}$(C$^{18}$O) with
\begin{eqnarray}
  \tau_{18}  =  -\ln(1-{\rm T}_{mb}({\rm C}^{18}{\rm O})/(5.27/(\exp(5.27/{\rm T}_{ex})-1)-0.87)). \nonumber \\  
\end{eqnarray}
Maps of the resulting C$^{18}$O opacity show that the line is
optically thin for all velocity components and reaches its maximum
value of $\tau_{18} \sim$0.4 only in the DR21 ridge for the -4 km
s$^{-1}$ component.  \\

\noindent{\bf H$_2$ column density from $^{12}$CO 1$\to$0}\\
\noindent The $^{12}$CO 1$\to$0 line is optically thick so that it should a
priori not be a good tracer for the total column density. However,
empirical studies (Strong et al. \cite{strong1988}, Bolatto et al.
\cite{bolatto2013}) show that the line still can trace
the total column density/mass of a molecular cloud
reasonably well with a conversion factor of 2--2.3$\times$10$^{20}$ from line integrated CO main-beam
brightness temperature into H$_2$ column density. Here, we adopt  a
value of 2$\times$10$^{20}$ K$^{-1}$ (km/s)$^{-1}$ cm$^{-2}$. \\

\noindent{\bf H$_2$ column density from CS 2$\to$1}\\
\noindent \noindent The values for the CS 2$\to$1 transition are
$\nu$=98.0 GHz, h$\nu$/k=4.70 K, $\mu$=1.95 Debye, $J_u$=2, and
$J(2.7K)$=1.00. The factor $f(T_{ex})$ is 16.1, 9.48, 7.35, 
7.00, and 7.41$\times$10$^{12}$ K$^{-1}$ (km/s)$^{-1}$ cm$^{-2}$ for 5, 7, 10, 15, and 20 K.
We assume optically thin CS emission and an abundance [CS]/[H$_2$] of
4$\times$10$^{-10}$ (Fechtenbaum et al., in prep.). \\

\noindent{\bf H$_2$ column density from N$_2$H$^+$ 1$\to$0}\\
\noindent The values for the N$_2$H$^+$ 1$\to$0 transition are
$\nu$=93.17 GHz, h$\nu$/k=4.47 K, $\mu$=3.40 Debye, $J_u$=1, and
$J(2.7K)$=1.06. The factor $f(T_{ex})$ is 3.62, 2.44, 2.10, 2.16, and
2.39$\times$10$^{12}$ K$^{-1}$ (km/s)$^{-1}$ cm$^{-2}$ for 5, 7, 10, 15, and 20
K.  The total beam averaged N$_2$H$^+$ column density is calculated
using the line-integrated emission from the F=(2,1) J=(1,0) component
at a frequency of 93.1737767 GHz and a HFS intensity ratio of 7/27 for that component.
We assume an abundance [N$_2$H$^+$]/[H$_2$] of 5$\times$10$^{-10}$. \\

We emphasise that all H$_2$ column-density determinations have a large
uncertainty, mainly arising from the abundance. The [CO]/[H$_2$]
abundance for the CO isotopolgues is probably better constrained
because it has been well studied in the literature.  However, the
variation in excitation temperature along the line of sight is more
significant here because the CO lines trace a larger range in
densities and temperatures than the high-density tracers CS and
N$_2$H$^+$. For the latter, the constant excitation temperature
assumption is most likely more valid, but the abundances of CS and
N$_2$H$^+$ are less well constrained, in particular for massive clumps
and cores.  In summary, we consider the H$_2$ column densities from CO
to be correct within a factor of two and the ones from CS and
N$_2$H$^+$ within a factor of a few.

\section{Individual molecular line PDFs }

In the main body of the paper, we used a normalization for the PDF
with respect to the size, i.e. pixel number of the dust map. We here
show the PDFs obtained with a normalization $\int_{-\infty}^{+\infty}
p_\eta \rm d \eta = \int_0^{+\infty} p_N\, dN=1$. As such, the PDFs
are consistent with what is shown in the literature (see Schneider et
al. 2015a and references therein) and the number of pixels defining
the PDF can be extracted. Figure~C.1 shows the resulting PDFs.

\begin{figure*}[!htpb]  
\begin{centering}  
\vspace{0cm}\includegraphics [width=8cm, angle={90}]{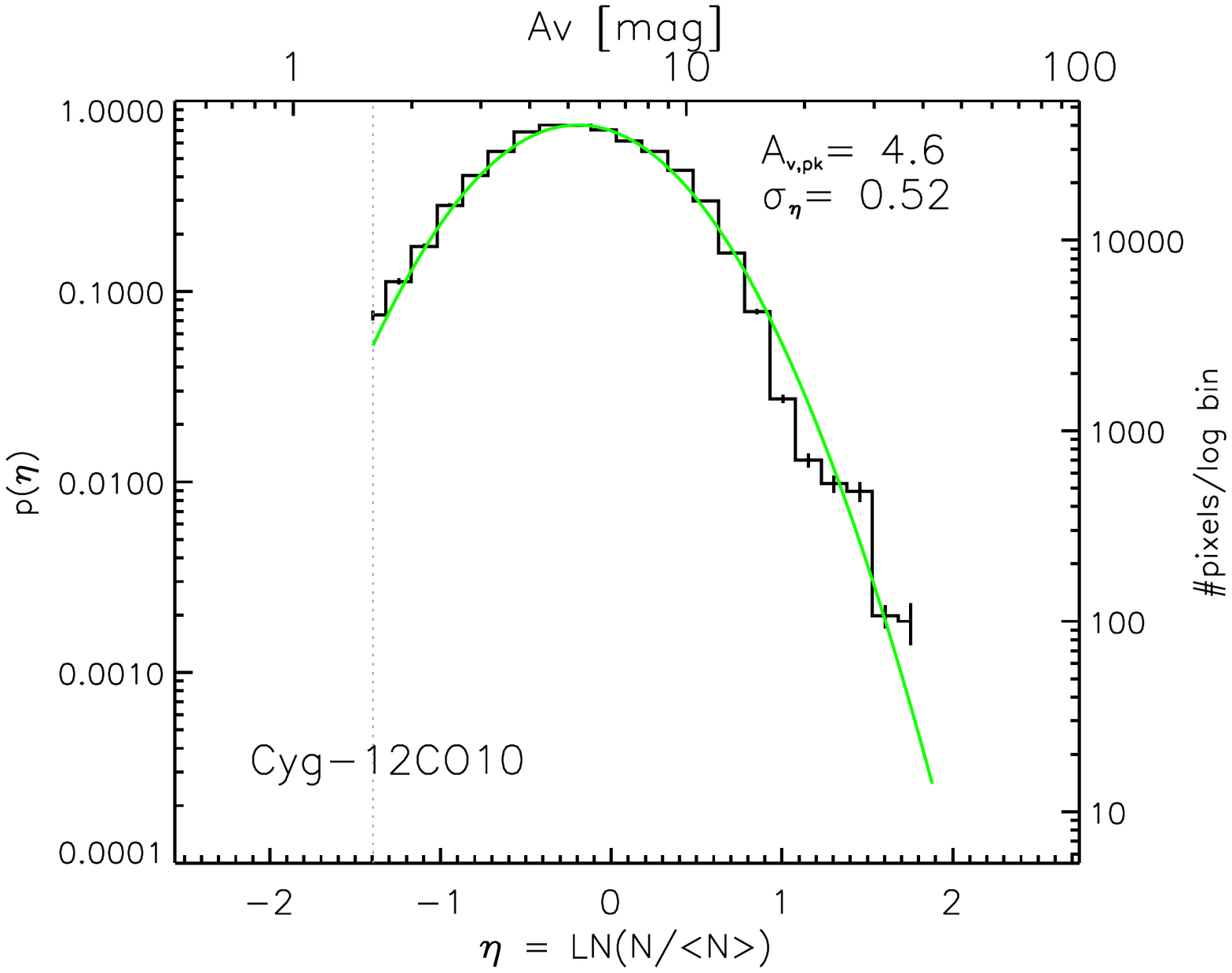}  
\hspace{-2.5cm}\includegraphics [width=8cm, angle={90}]{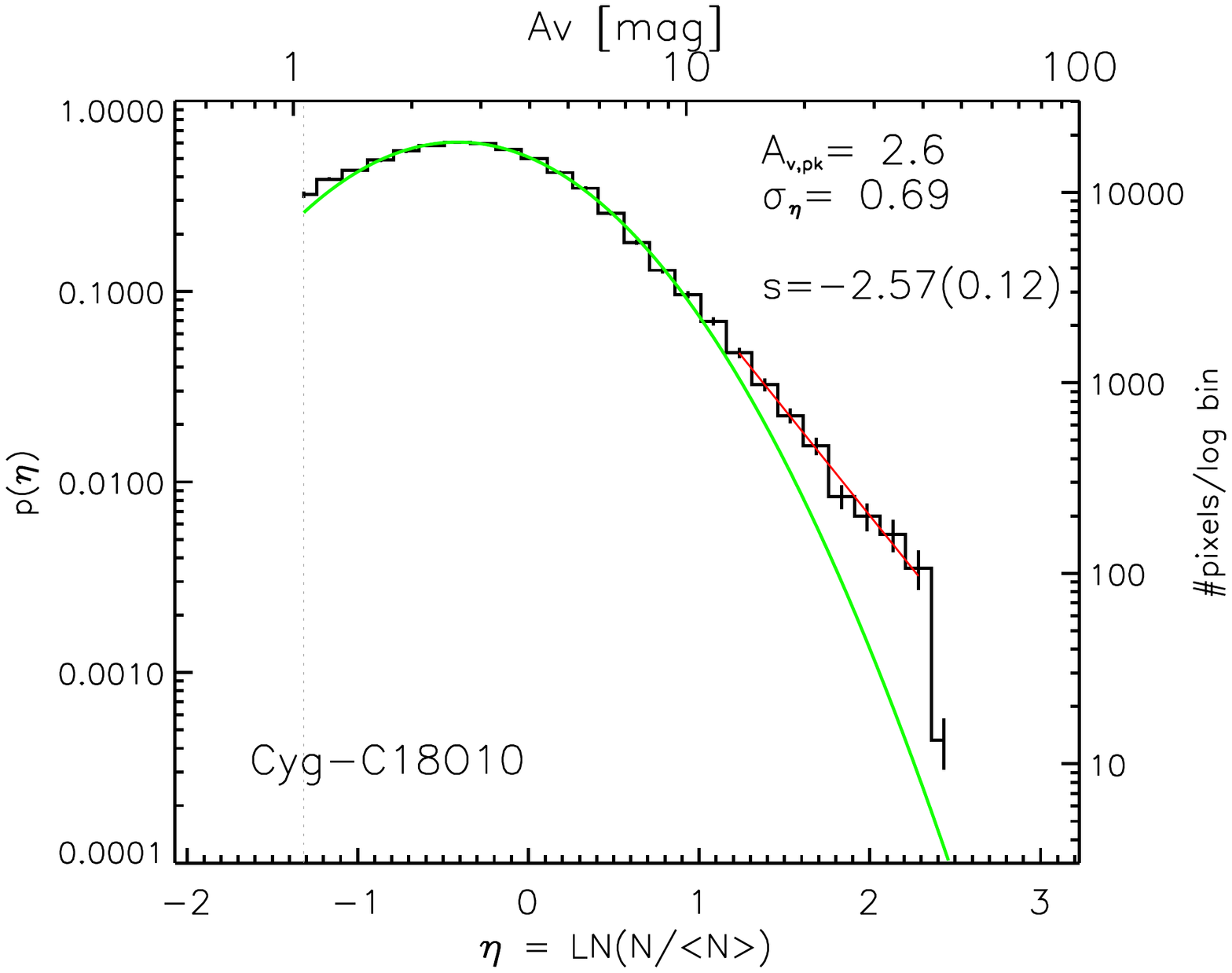}  
 
\vspace{-1.0cm}\includegraphics [width=8cm, angle={90}]{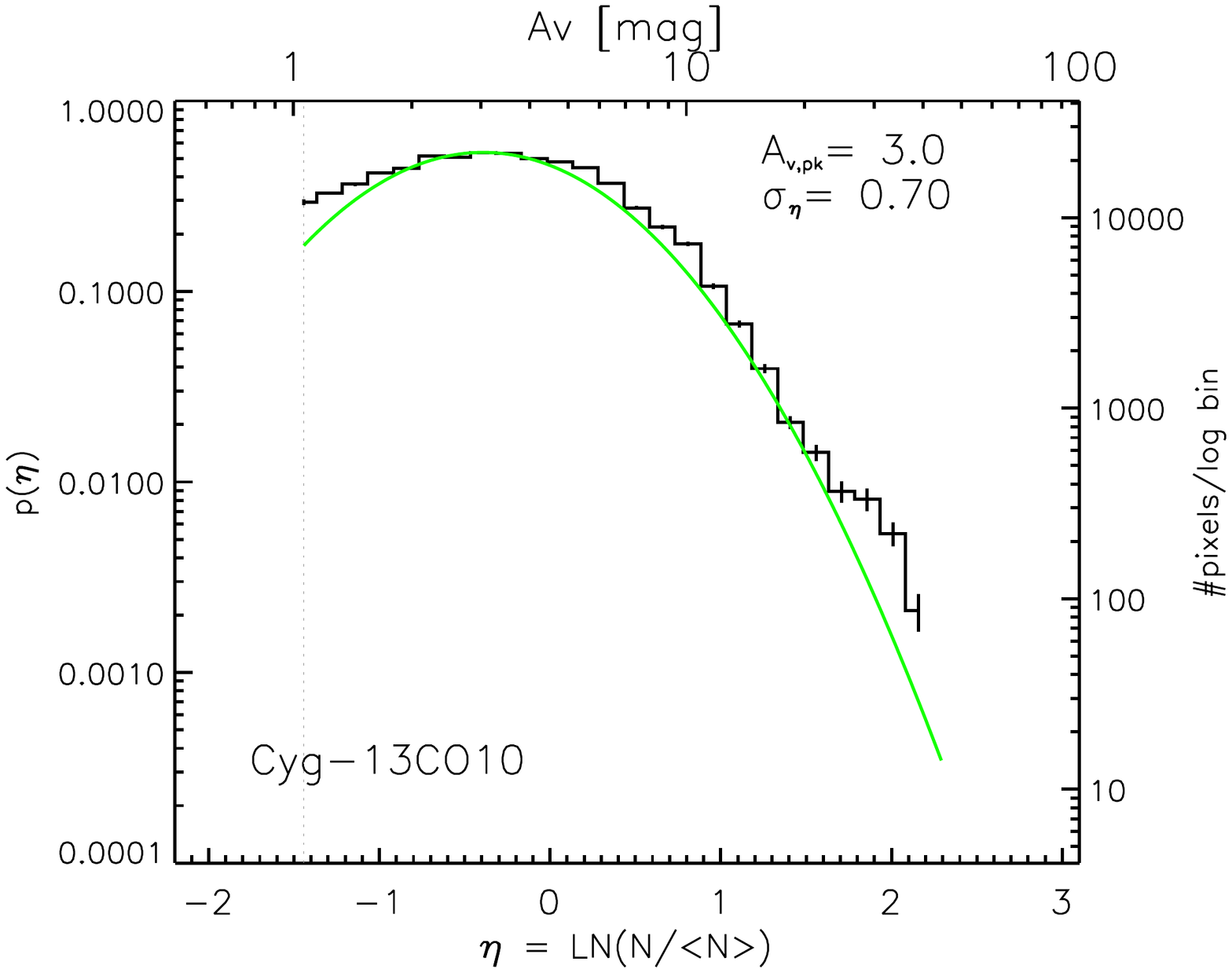}  
\hspace{-2.5cm}

\vspace{-1.0cm}\includegraphics [width=8cm, angle={90}]{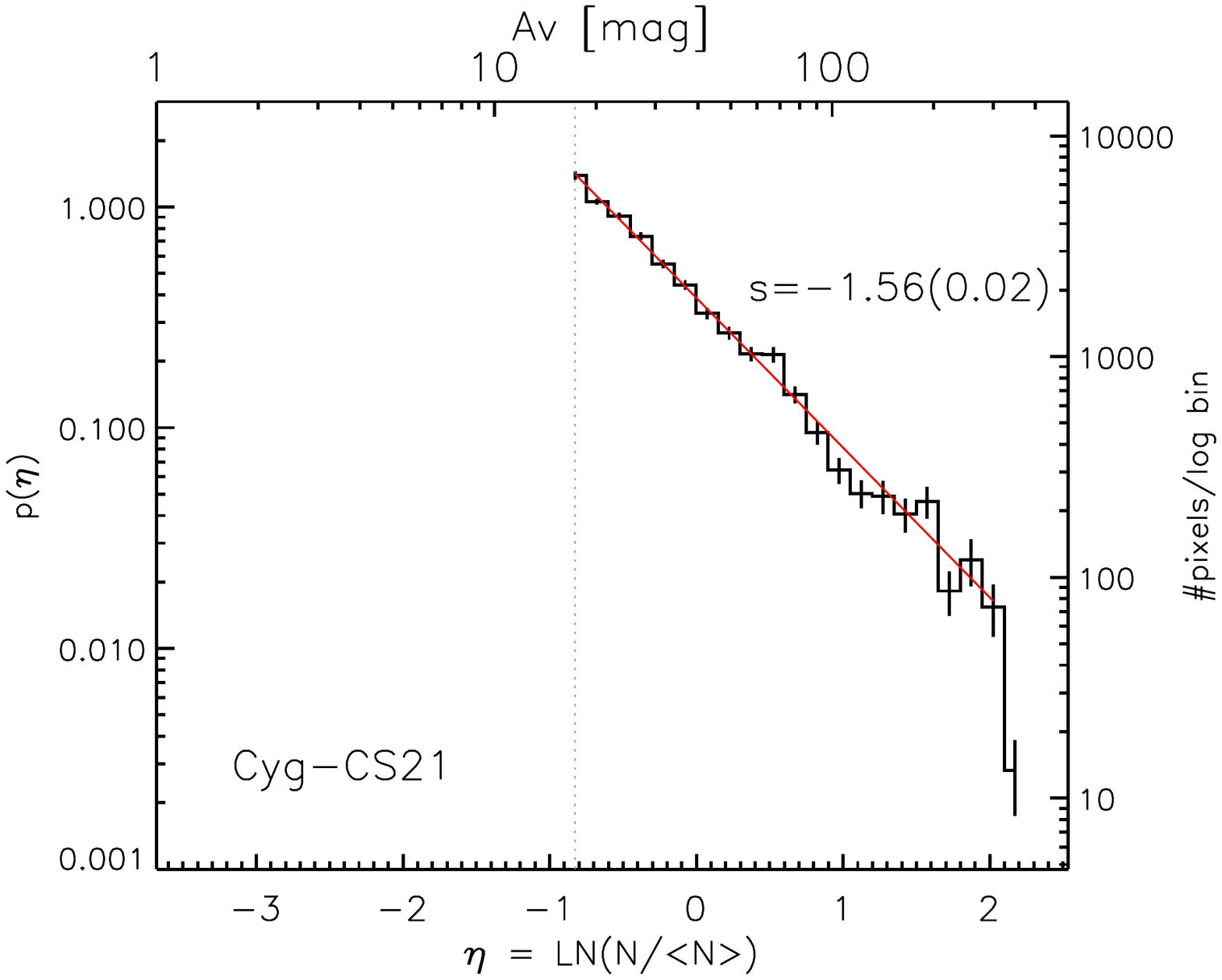}  
\hspace{-2.5cm}\includegraphics [width=8cm, angle={90}]{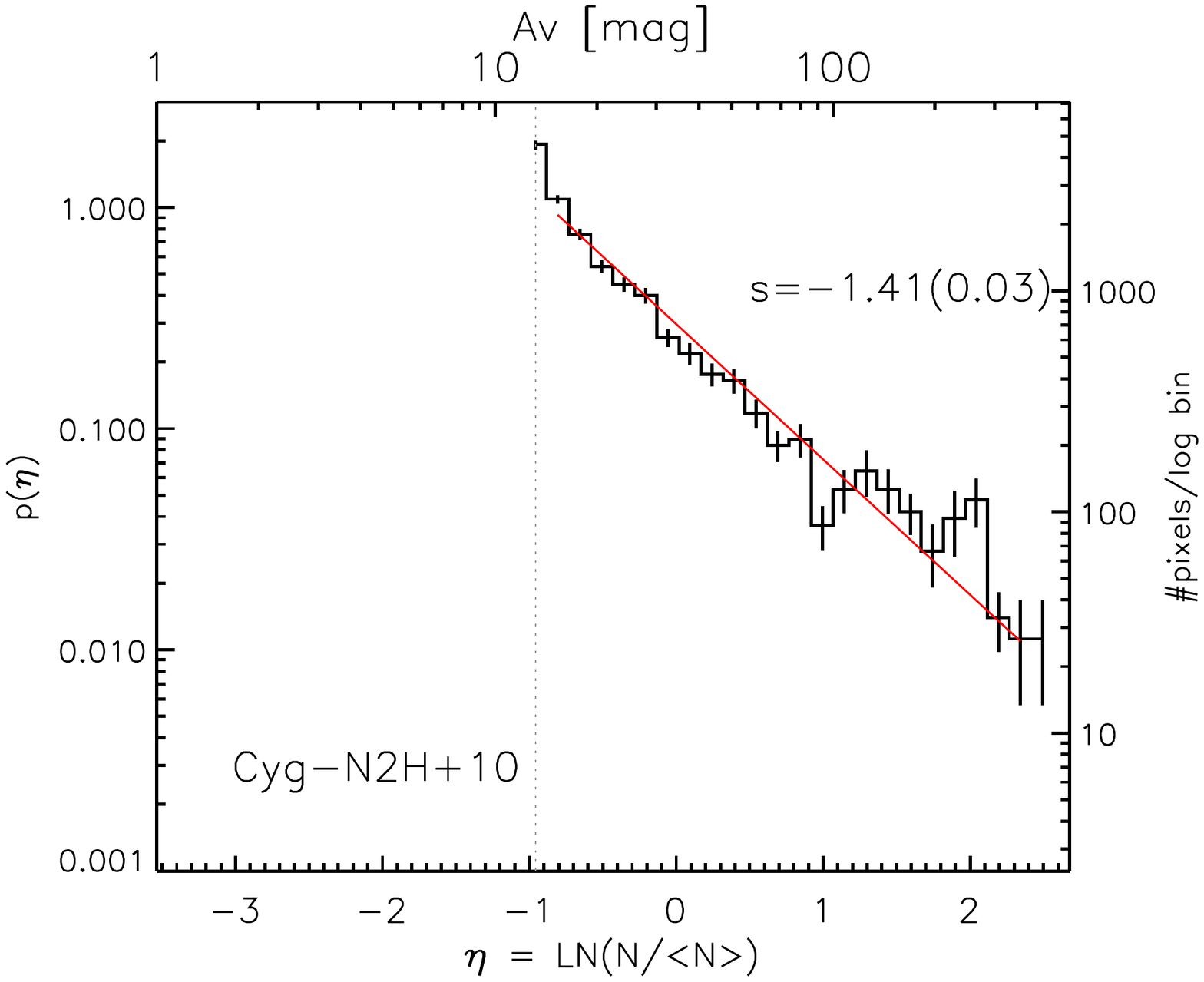}  

\caption[] {PDFs derived from the molecular line maps. The left y-axis
  gives the normalized probability $p(\eta)$, the right y-axis the
  number of pixels per log bin. The upper x-axis is the visual
  extinction and the lower x-axis the logarithm of the normalized
  column density.  For the CO lines, we fitted a lognormal to the low
  column-density range, indicated by a green curve. The red line
  shows the power-law fit to the high column-density end for
  C$^{18}$O, CS, and N$_2$H$^+$. The respective slopes $s$ (and errors) are given
  in the panels. For the CO lines, the peak and width of the PDF is indicated with \av$_{peak}$ and $\sigma$.}
\label{pdf-mol}  
\end{centering} 
\end{figure*}

\section{Relation between PDF slope and radial density distribution for different geometries}

We define the PDF $p_N$ as the surface fraction with gas of column
density $N$ and the PDF of the natural log of the column density
density $p_\eta$ with $\eta=\ln(N/\langle N \rangle)$, the relation
between the two is $p_\eta=N p_N$. The definition of $p_N$ gives $p_N
\propto dA/dN$. \\

\noindent {\bf Spherical geometry} \\  
\noindent In the case of spherical geometry, the density radial
profile is $\rho(r) \propto r^{-\alpha}$, hence $N \propto \rho \, r
\propto r^{-\alpha+1}$ ($dN \propto r^{-\alpha}$). Hence, with the
area $A \propto r^2$ ($dA \propto r$)
\begin{eqnarray}
p_N &\propto& dA/dN \propto r/r^{-\alpha} \propto N^{(1+\alpha)/(1-\alpha)}\cr
p_\eta &=& N p_N \propto N^{2/(1-\alpha)}
.\end{eqnarray}
\noindent {\bf Cylindrical geometry} \\
\noindent In the case of cylindrical geometry, the density radial profile from the axis of
the cylinder is $\rho(r_c)\propto r_c^{-\alpha}$, hence $N \propto
\rho \, r_c \propto r_c^{-\alpha+1}$. However in this case $A\propto
z \, \times \, r_c \propto r_c$, hence
\begin{eqnarray}
p_N &\propto& dA/dN \propto 1/r_c^{-\alpha} \propto
N^{\alpha/(1-\alpha)}\cr p_\eta &=& N p_N \propto N^{1/(1-\alpha)}
.\end{eqnarray}
Note that similar calculations can be found in Federrath \& Klessen (\cite{fed2013}), Fischera et al. (\cite{fischera2014}), and
Myers (\cite{myers2015}). 


\begin{thebibliography}{}   
  
\bibitem[2014]{catarina2014}     
Alves de Oliveira, C., Schneider, N., Merin, B., et al., 2014, A\&A568, 98 

\bibitem[2001]{bergin2001}   
Bergin, E.~A., Ciardi, D.R., Lada, C., 2001, ApJ, 557, 209
  
\bibitem[1995]{blake1995}
Blake, G.A., Sandell, G., van Dishoeck, E.F., et al., 1995, ApJ, 441, 689
   
\bibitem[1978]{bohlin1978}   
Bohlin, R.C., Savage, B.D., Drake, J.F., 1978, ApJ 224, 132  
 
\bibitem[2010]{bontemps2010}   
Bontemps, S., Motte, F., Csengeri, T., et al., 2010, A\&A, 524, 18  
 
\bibitem[2013]{bolatto2013}   
Bolatto, A.D., Wolfire, M., Leroy, A.K., 2013, ARA\&A, 51, 207 
  
\bibitem[2013]{burk2013}     
Burkhart, B., Ossenkopf, V., Lazarian, A., Stutzki, J., 2013, ApJ, 771, 122 
  
\bibitem[2013]{carlhoff2013}   
Carlhoff, P., Schilke, P., Nguyen-Luong, Q., et al., 2013, A\&A, 560, 24

\bibitem[2002]{caselli2002}   
Caselli, P., Benson, P.J. Myers, P.C., Tafalla, M., 2002, ApJ, 572, 238
  
\bibitem[2011]{csengeri2011}   
Csengeri, T., Bontemps, S., Schneider, N., et al., 2011, A\&A, 527, 135  
  
\bibitem[2005]{daniel2005}   
Daniel, F., Dubernet, M.-L., Meuwly, M., et al., 2005, MNRAS, 363, 1083

\bibitem[2008]{drew2008}   
Drew, J.E., Greimel, R., Irwin, M.J., Sale, S.E., 2008, MNRAS, 386, 1761

\bibitem[2013]{duarte2013}   
Duarte-Cabral, A., Bontemps, S., Motte, F., et al., 2013, A\&A, 558, 125
   
\bibitem[2008]{fed2008}   
Federrath, C., Klessen, R.S., Schmidt, W., 2008, ApJ, 688, L79  
  
\bibitem[2010]{fed2010}   
Federrath, C., Roman-Duval, J., Klessen, R.J., et al., 2010, A\&A, 512, 81  
  
\bibitem[2012]{fed2012}   
Federrath, C., Klessen, R. S., 2012, ApJ, 761, 156 
 
\bibitem[2013]{fed2013}   
Federrath, C., Klessen, R. S., 2013, ApJ, 763, 51  

\bibitem[2014]{fischera2014}   
Fischera, J., 2014, A\&A, 571, 95

\bibitem[1999]{flower1999}
Flower, D.R., 1999, MNRAS, 305, 651
  
\bibitem[2012]{fontani2012}   
Fontani, F., Giannetti, A., Beltran, M.T., et al., 2012, MNRAS, 423, 2342  
  
\bibitem[1982]{freking1982}   
Frerking, M.A., Langer, W.D., Wilson, R.W., 1982, ApJ, 262, 590 
  
\bibitem[2010]{froebrich2010}   
Froebrich, D., Rowles, J., 2010, MNRAS, 406, 1350  

\bibitem[2012]{gianetti2012}   
Gianetti, A., Brand, J., Massi, F., et al., 2012, A\&A, 538, 41
  
\bibitem[2014]{giri2014}   
Girichidis, P., Konstandin, L., Whitworth, A.P., Klessen, R., et al., 2014, ApJ, 781, 91  
 
\bibitem[2008]{goldsmith2008}     
Goldsmith, P.F., Heyer, M., Narayanan, G., et al., 2008, ApJ, 680, 428    
 
\bibitem[2009]{good2009}  
Goodman, A.A., Pineda, J.E., Schnee, S.L., 2009, ApJ, 692, 91  
  
\bibitem[2010]{griffin2010}  
Griffin, M., Abergel, A., Abreau, S., et al., 2010, A\&A, 518   
  
\bibitem[2012]{guarcello2012}  
Guarcello, M.G., Wright, N.J., Drake, J.J., et al., 2012, ApJS, 202, 19   
  
\bibitem[2008]{hennebelle2008}   
Hennebelle, P., Chabrier, G., 2008, ApJ, 684, 395  
  
\bibitem[2012]{hennemann2012}   
Hennemann, M., Motte, F., Schneider, N., et al., 2012, A\&A, 543, L3  
  
\bibitem[2011]{hill2011}   
Hill, T., Motte F., Didelon P., et al., 2011, A\&A, 533, 94   
 
\bibitem[2013]{hopkins2013}   
Hopkins, P.F., 2013, MNRAS, 423, 2037

\bibitem[2009]{kai2009}  
Kainulainen, J., Beuther, H., Henning, T., \& Plume, R., 2009, A\&A, 508, L35  
  
\bibitem[2013]{kirk2013}  
Kirk, H., Myers, P.C., Bourke, T.L., et al., 2013, ApJ, 766, 115
  
\bibitem[2000]{klessen2000}  
Klessen, R.~S., 2000, ApJ, 535, 869   
   
\bibitem[2015]{vera2015}  
K\"onyves, V., Andr\'e P., Men'shchikov A., et al., 2015, A\&A, 584, 91
  
\bibitem[2005]{krumholz2005}  
Krumholz, M., McKee, C.F., 2005, ApJ, 630, 250
  
\bibitem[1990]{langer1990}   
Langer, W.D., Penzias, A.A., 1990, ApJ, 357, 477

\bibitem[1969]{larson1969}    
Larson, R.B., 1969, MNRAS, 145, 271
   
\bibitem[2015]{li2015}   
Li, J., Wang, J., Zhu, Q., et al., 2015, ApJ, 802, 40
  
\bibitem[2009]{lo2009}   
Lo, N., Cunningham, M.R., Jones, P.A., et al., 2009, MNRAS, 395, 1021
    
\bibitem[2008]{lombardi2008}  
Lombardi, M., Lada, C., Alves, J., 2008, A\&A, 489, 143  
  
\bibitem[2015]{mangum2015}
Mangum, J.G., Shirley, Y.L., 2015, PASP, 127, 266

\bibitem[2007]{motte2007}   
Motte, F., Bontemps, S., Schilke P., et al., 2007, A\&A,  476, 1243  
  
\bibitem[2010]{motte2010}   
Motte, F.,  Zavagno A., Bontemps S., et al., 2010, A\&A 518, L77   

\bibitem[2015]{myers2015}   
Myers, P.C., 2015, ApJ, 806, 226
    
\bibitem[2015]{neufeld2015}   
Neufeld, D.A., Godard, B., Gerin, M., et al., 2015, A\&A, 577, 49

\bibitem[2011]{padoan2011}   
Padoan, P., Nordlund, A., 2011, ApJ, 741, 22
   
\bibitem[2013]{pedro2013}   
Palmeirirm, P., Andr\'e P., Kirk, J.,  et al., 2013, A\&A, 550, 38

\bibitem[1969]{penston1969}    
Penston, M.V., 1969, MNRAS, 144, 425

\bibitem[2013]{peretto2013}    
Peretto, N., Fuller, G.A., Duarte-Cabral, A., et al., 2013, A\&A, 555, 112

\bibitem[2008]{pineda2008}   
Pineda, J.E., Caselli, P., Goodman, A., 2008, ApJ, 679, 481

\bibitem[2010]{pineda2010}   
Pineda, J.L., Goldsmith, P.F., Chapman N., et al., 2010, ApJ, 721, 686
    
\bibitem[2003]{pigorov2003}   
Pigorov, L., Zinchenko, I., Caselli, P., et al., 2003, A\&A, 405, 639
  
\bibitem[2010]{poglitsch2010}  
Poglitsch, A., Waelkens, C., Geis, N., et al., 2010, A\&A 518    

\bibitem[2008]{reipurth2008}  
Reipurth, B., Schneider, N., 2008, Handbook of star forming regions, Vol. I, ASP Publ., Ed. B. Reipurth, p.36

\bibitem[2013]{ripple2013}  
Ripple, F., Heyer, M.H., Gutermuth, R., et al., 2013, MNRAS, 431, 1296 
  
\bibitem[2013]{russeil2013}    
Russeil, D., Schneider, N., Anderson, L., et al., 2013, A\&A, 554, 42  
  
\bibitem[2012]{rygl2012} 
Rygl, K.L.J., Brunthaler, A., Sanna, A., et al., 2012, A\&A, 539, A79 

\bibitem[2009]{sale2009}   
Sale, S.E., Drew, J.E., Unruh, Y.C., et al., 2009, MNRAS, 392, 497   

\bibitem[2006]{schneider2006}   
Schneider, N., Bontemps, S., Simon, R., et al., 2006, A\&A, 458, 855   
 
\bibitem[2007]{schneider2007}   
Schneider, N., Simon, R., Bontemps, S, et al., 2007, A\&A, 474, 873   
  
\bibitem[2010]{schneider2010}   
Schneider, N.,  Csengeri T., Bontemps S., Motte F., Simon R., Hennebelle P., Federrath C., 
Klessen R., 2010, A\&A, 520, 49 
 
\bibitem[2011]{schneider2011}   
Schneider, N., Bontemps, S., Simon, R., et al., 2011, A\&A,  529, 1  
  
\bibitem[2012]{schneider2012}   
Schneider, N., Csengeri, T., Hennemann, M., et al., 2012, A\&A, 540, L11  
 
\bibitem[2013]{schneider2013}   
Schneider, N., Andr\'e, Ph., K\"onyves, V., Bontemps, S., et al., 2013, ApJ, 766, L17

\bibitem[2015a]{schneider2015a}   
Schneider, N., Ossenkopf, V., Csengeri, T., et al., 2015a, A\&A, 575, 79

\bibitem[2015b]{schneider2015b}   
Schneider, N., Csengeri, T., Klessen, R.S, et al., 2015b, A\&A, 578, 29

\bibitem[2015c]{schneider2015c}   
Schneider, N., Bontemps, S., Girichidis, P., et al., 2015c, MNRAS, 453, L41

\bibitem[2015d]{schneider2015d}   
Schneider, N., Bontemps, S., Motte, F., 2015d, A\&A, in prep. 

\bibitem[2015]{shirley2015}
Shirley, Y.L., 2015, PASP, 127, 299

\bibitem[2011]{shetty2011}
Shetty, R., Glover, S.C., Dullemond, C.P., Klessen, R.S., 2011, MNRAS, 412, 1686
  
\bibitem[1977]{shu1977}
Shu., F., 1977, ApJ, 214, 488

\bibitem[1988]{strong1988} 
Strong, J., Bloemen, J., Dame, T., et al. 1988, A\&A, 207, 1 

\bibitem[1990]{stutzki1990} 
Stutzki, J., G\"usten, R., 1990, ApJ, 356, 513

\bibitem[2015]{stutz2015} 
Stutz, A.M., Kainulainen, J., 2015, A\&A, 577, L6
    
\bibitem[2015]{laszlo2015} 
Sz\"ucs, L., Glover, S.C.O., Klessen, R.S., 2015, MNRAS, in prep. 

\bibitem[2002]{tafalla2002}   
Tafalla, M., Myers, P.C., Caselli, P., Walmsley, C.M., 2002, ApJ, 569, 815

\bibitem[2006]{tafalla2006}   
Tafalla, M., Santiago-Garcia, J., Myers, P.C., et al., 2006, A\&A, 455, 577 
   
\bibitem[2015]{toci2015}   
Toci, C., Galli, D., 2015, MNRAS, 446, 2118

\bibitem[2014]{tremblin2014}   
Tremblin, P., Schneider, N., Minier, V., et al., 2014, A\&A, 564, 106
  
\bibitem[2001]{vaz2001}    
V{\'a}zquez-Semadeni, E., Garcia, N., 2001, ApJ, 557, 727    
       
\bibitem[1988]{ewine1988}   
Van Dishoeck, E.F., \& Black, 1988, ApJ, 334, 771  
  
\bibitem[2014]{ward2014}     
Ward, R.L., Wadsley, J., Sills, A., 2014, MNRAS, 445, 1575

\bibitem[1994]{wilson1994}     
Wilson, T.L., Rood, R., 1994, ARA\&A, 32, 191
  
\bibitem[1985]{whitworth1985}     
Whitworth, A., Summers et al., 1985, MNRAS, 214, 1 
   
\bibitem[2008]{wong2008}   
Wong, T., Ladd, E.F., Brisbin, D., et al., 2008, MNRAS, 386, 1069   

\bibitem[2010]{wright2010}   
Wright, N.J., Drake, J.J., Drew, J.E., Vink, J.S., 2010, ApJ, 713, 871   

\end{thebibliography}
\end{document}